\documentclass[twocolumn,showpacs,amsmath,amssymb,superscriptaddress,aps]{revtex4-2}

\usepackage{graphicx}% Include figure files
\usepackage{dcolumn}% Align table columns on decimal point
\usepackage{bm}% bold math
\usepackage{amsmath,amssymb}
\usepackage{xcolor}
\usepackage{enumitem}

\begin{document}

\preprint{APS/123-QED}

\title{Improvement of graviton mass constraints using GRAVITY's detection of Schwarzschild precession in the orbit of S2 star around the Galactic Center}

\author{Predrag Jovanovi\'{c}}
\affiliation{Astronomical Observatory, Volgina 7, P.O. Box 74, 11060 Belgrade, Serbia}

\author{Vesna Borka Jovanovi\'{c}}
\affiliation{Department of Theoretical Physics and Condensed Matter Physics (020), Vin\v{c}a Institute of Nuclear Sciences - National Institute of the Republic of Serbia, University of Belgrade, P.O. Box 522, 11001 Belgrade, Serbia}

\author{Du\v{s}ko Borka}
\email[Corresponding author:]{dusborka@vinca.rs}
\affiliation{Department of Theoretical Physics and Condensed Matter Physics (020), Vin\v{c}a Institute of Nuclear Sciences - National Institute of the Republic of Serbia, University of Belgrade, P.O. Box 522, 11001 Belgrade, Serbia}

\author{Alexander F. Zakharov}
\affiliation{Bogoliubov Laboratory for Theoretical Physics, JINR, 141980 Dubna, Russia}

\date{\today}

\begin{abstract}
Here we study possible improvements of the existing constraints on the upper bound of graviton mass by the analysis of the stellar orbits around the supermassive black hole (SMBH) at the Galactic Center (GC) in the framework of Yukawa gravity. A motivation for this study is a recent detection of Schwarzschild precession in the orbit of S2 star around the SMBH at the GC by the GRAVITY Collaboration. The authors indicated that the orbital precession of the S2 star is close to the General Relativity (GR) prediction, but with possible small deviation from it, and parametrized this effect by introducing an ad hoc factor in the parameterized post-Newtonian (PPN) equations of motion. Here we use the value of this factor presented by GRAVITY in order to perform two-body simulations of the stellar orbits in massive gravity using equations of motion in the modified PPN formalism, as well as to constrain the range of massive interaction $\Lambda$. From the obtained values of $\Lambda$, and assuming that it corresponds to the Compton wavelength of graviton, we then calculated new estimates for the upper bound of graviton mass which are found to be independent, but consistent with the LIGO's estimate of graviton mass from the first gravitational wave (GW) signal GW150914 (later this graviton mass estimation was significantly improved with consequent observations of GW events). We also performed calculations including numerical simulations in order to constrain the bounds on graviton mass in the case of a small deviation of the stellar orbits from the corresponding GR predictions and showed that our method could further improve previous estimates for upper bounds on the graviton mass. It is also demonstrated that such an analysis of the observed orbits of S-stars around the GC in the frame of the Yukawa gravity represents a tool for constraining the upper bound for the graviton mass, as well as for probing the predictions of GR or other gravity theories.
\end{abstract}

\pacs{alternative theories of gravity, supermassive black hole, stellar dynamics}
\maketitle

\section{Introduction}

There are a number of modified gravity theories which have been proposed to explain cosmological and astrophysical data at different scales without introducing dark energy and dark matter (see, e.g., \cite{fisc99,kope04,clif06,capo11a,capo11b,noji11,capo12a,clif12,noji17,salu21,li17} for reviews and references therein). Several of them have been proposed as possible extensions of Einstein's theory of gravity \cite{fisc99,capo11a,capo11b,capo12a}. Also, the theories of massive gravity have attracted a lot of attention (see e.g. \cite{gers04,gers06,ruba08,babi10,rham11,rham14,rham17} and references therein). Different modified gravity theories, including those with massive graviton, are also widely used for studying the black hole physics. For example, the black hole shadows of rotating black hole solutions in $f(R)$ gravity with nonlinear electrodynamics were investigated in \cite{sun23}, while the thermodynamics and entropy of black hole solutions for massive gravity were studied in \cite{hu20,zhou20}.

Massive gravity theories are the theories beyond Einstein's gravity theory with massless graviton, and some cosmologists have proposed the idea of a massive graviton to modify general relativity. In these theories graviton has some small, nonzero mass $m_g$ because gravity is propagated by a massive field. The first version of a massive gravity theory was proposed by Fierz and Pauli \cite{fier39}, however, later pathologies were found in this approach, for instance, Boulware and Deser found ghosts in this approach \cite{boul72} and a massive gravity theory was not very attractive for a while. However, later a way to create a massive gravity theory without ghosts was proposed \cite{rham11,rham14,rham17}.

Currently, a theory of massive gravity is treated as a reasonable alternative for general relativity (GR) and in the first publication on the discoveries of gravitational waves and binary black holes, the LIGO and Virgo collaborations considered a theory of massive gravity as a suitable approach and the authors reported their estimate for graviton mass $m_g < 1.2\times 10^{-22}~$eV from analysis of the first gravitational wave detection \cite{LIGO_16}. Analyzing observational data of gravitational waves from three observational runs LIGO -- Virgo -- KAGRA collaborations significantly improved this constraint up to $m_g < 1.27\times 10^{-23}~$eV \cite{LIGO_21}. Different experimental limits on the mass of graviton are given in \cite{zyla20} and references therein. In a review by de Rham \cite{rham14} the discussion \textcolor{red}{of} different theories, such as massive gravity models from extra dimensions, ghost-free massive gravity, as well as Lorentz-violating and non-local massive gravity theory, is provided. In paper \cite{miao19} the authors constrained a graviton mass in a dynamic regime with binary pulsars. Some massive gravity theories belong to a class of Yukawa gravity models. In these theories, bounds on the graviton come from the exponential decay in the Yukawa potential which switches gravity off at the graviton's Compton wavelength \cite{rham14}. Current state-of-art of Yukawa-like potentials are given in Table 1 of Ref. \cite{capo21}. The authors reported various Extended Theories of Gravity where Yukawa-like corrections in the post-Newtonian limit are the general feature.

The main characteristic of Yukawa-like potentials are a presence of decreasing exponential terms \cite{sand84,talm88,sere06,card11,iori07,iori08}. The Yukawa-like modification of the Newtonian gravitational potential is given by:
\begin{equation}
\Phi\left(r\right)=-\dfrac{GM}{(1+\delta)r}\left({1+\delta e^{-\dfrac{r}{\Lambda}}}\right),
\label{yukawa}
\end{equation}
where $\Lambda$ is the range of interaction which depends on the typical scale of a gravitational system, while $\delta$ is the strength of interaction.

The Yukawa term is extensively studied at short ranges (see \cite{adel09} and references therein), as well as at long ranges in the case of clusters of galaxies \cite{capo07b,capo09b} and rotation curves of spiral galaxies \cite{card11}. Other studies of long-range Yukawa term investigations can be found in \cite{whit01,amen04,reyn05,seal05,sere06,moff05,moff06}. Besides, a number of authors tried to explain observations that have recently emerged at different astrophysical scales using Yukawa-like gravity framework \cite{rham17,will18,mart21,beni22b,miao19,dong22}. In paper \cite{rham17}, the different graviton mass bounds obtained from massive potentials, Yukawa like and non-Yukawa like, at different scales (Solar System, galactic clusters and weak lensing) are reviewed and compared (with bound from gravitational waves GW150914). In Ref. \cite{will18} the Solar System data in case of Yukawa form of gravitation potential are analyzed and used to obtain the bounds on graviton mass. In the reference \cite{mart21} the authors studied the orbital precession of S2 star by modelling its orbit with geodesics. In \cite{beni22b} the Yukawa-like gravity was also investigated at Solar System scale, while in \cite{dong22} the authors presented analysis of Yukawa gravity parameters in the case of pulsars around Sgr A$^{*}$.

The compact bright radio source Sgr A$^\ast$ is located at the GC and S-stars are the bright stars which move around it \cite{ghez00,scho02,ghez08,gill09a,gill09b,genz10,meye12,gill17,hees17,chu18,abut18,abut19,do19,amor19,hees20,abut20}. It is shown that total mass of the GC mostly consists of the SMBH (with mass around $4.3\times 10^6 M_\odot$), and potentially of a much lesser content of a mass formed by a bulk mass distribution of stellar cluster, interstellar gas and probably dark matter \cite{abut22}. More detailed analysis can be found in key studies on the distributed mass component \cite{heib22,jian85,rubi01,grav23,naoz20,will23,evan23}. Except distributed mass components, one of the possible scenarios is the potential presence of companion black hole \cite{grav23,naoz20,evan23,will23}. 

The orbits of S-stars around Sgr A$^\ast$ are monitored for about 30 years by the New Technology Telescope and Very Large Telescope (NTT/VLT) in Chile \cite{gill09a,gill09b} and by Keck telescopes in Hawaii \cite{ghez08}. Also, Saida et al. reported observational data obtained with the SUBARU telescope by 2018. They have observed S2 star for more than 10 nights with the SUBARU telescope \cite{said19}. Recently, VLT units also started to operate as the GRAVITY interferometer. These scientific groups performed the precise astrometric observations of many S-stars \cite{ghez00,scho02,ghez08,gill09a,gill09b,genz10,gill17,hees17,chu18,abut18,abut19,do19,amor19,hees20,abut20,abut22}. Also, there is a number of recent analyses of different S-star orbits using available observational data performed by these two groups (see e.g. \cite{kali20,lalr21,lalr22,doku15,dela18,dadd21,bork21a,bork22a,beni22a}).

In 2020, the GRAVITY Collaboration detected the orbital precession of the S2 star around the supermassive black hole (SMBH) at the Galactic Center and showed that it is close to the GR prediction, as well as that a possible small deviation from it cannot be presently ruled out \cite{abut20}. Also, in paper \cite{dadd21} the author performs data analysis in the framework of Yukawa gravity model by solving geodesic equation, and concluded that the orbital precessions of the S2, S38 and S55 stars are close to the corresponding prediction of GR for these stars.

The aim of this paper is to constrain the upper bound for the graviton mass and to probe the predictions of GR by analysis of the stellar orbits around GC in the framework of Yukawa gravity, and taking into account the recent detection of Schwarzschild precession in the orbit of S2 star. We expect that future observations of bright stars will demonstrate similar evidences from
Schwarzschild precessions for other star orbits. For that purpose we studied the orbits of S-stars around the central SMBH of our Galaxy in the frame of Yukawa gravity using the modified PPN formalism \cite{clif08,alsi12,gain20a,gain20b}. We performed calculations for the same $f_{SP}$ values and their measured precision for all S-stars from Table 3 in \cite{gill17} except of S111, we suppose that $f_{SP}$ will be very near to GR value (we assumed that the above measurements are a confirmation of GR within 1 $\sigma$), and these are the values for the Schwarzschild precession detected for S2 by GRAVITY collaboration \cite{abut20,abut22} and for combination of a few stars: S2, S29, S38 and S55 also detected by GRAVITY collaboration \cite{abut22}. Also, we want to analyse what will happen in the future observations if a GR prediction will be confirmed with much higher accuracy, i.e. if value of $f_{SP}$ becomes much closer to 1, and when absolute error $\Delta f_{SP}$ becomes much smaller (currently $f_{SP}=1.10\pm0.19$, i.e. $\Delta f_{SP}$ = 0.19, \cite{abut20}; and the latest updated values: $f_{SP}=0.85\pm0.16$ and $f_{SP}=0.997\pm0.144$ \cite{abut22}). That is why we also analyse cases $f_{SP}=1.01\pm0.005$ and $f_{SP}=1.001\pm0.0005$. This research is continuation of our previous investigations of different Extended Gravity theories where we used astronomical data for different astrophysical systems: the S2 star orbit \cite{bork12,bork13,zakh14,capo14,bork16,zakh16a,zakh16b,zakh17a,zakh17b,zakh18a,zakh18b,dial19,bork19,jova21,bork21a,bork22a,zakh22a,zakh22b,jova23}, fundamental plane of elliptical galaxies \cite{bork16a,capo20,bork21b,bork23} and baryonic Tully-Fischer relation of spiral galaxies \cite{capo17}. Specifically, this research is closely related to our previous studies \cite{zakh16a,zakh18a,jova23}, but with much novel elements which are added in the present work. In paper \cite{zakh16a} we considered phenomenological consequences of massive gravity and showed that an analysis of bright star trajectories could bound the graviton mass. Using simulations of the S2 star orbit around the SMBH at the Galactic Center in Yukawa gravity and their comparisons with the NTT/VLT astrometric observations of S2 star \cite{gill09a} we get the constraints on the range of Yukawa interaction which showed that $\Lambda > 4.3\times 10^{11}$ km. Taking this value as the lower bound for the graviton Compton wavelength, we found that the corresponding most likely upper bound for graviton mass is $m_g < 2.9\times 10^{-21}$ eV.
In paper \cite{zakh18a} in contrast with our previous studies \cite{zakh16a}, we presented current constraints on parameters of Yukawa gravity and graviton mass, assuming the values for orbital precession in the case of all S-stars (Table 3 from \cite{gill17}) will be equal to the corresponding GR orbital precession for each star.
In paper \cite{jova23} we constrained the Yukawa gravity parameters from the observations of bright stars, but we did not calculate the upper bound of graviton mass. Our main goal was to study the possible influence of the strength of Yukawa interaction, i.e. the universal constant $\delta$, on the precessions of S-star orbits. We analyzed the S-star orbits assuming different strengths of Yukawa interaction $\delta$ and found that this parameter had strong influence on the range of Yukawa interaction $\Lambda$. 

The main goal of this study is to use the recent detection of Schwarzschild precession by the GRAVITY Collaboration in order to obtain the new estimates for graviton mass (we investigate if its slightly different value than that predicted by GR could be explained by a non-zero graviton mass). In this paper: 1) We study the possible improvements of the existing constraints on the upper bound of a graviton mass by the analysis of the stellar orbits around the SMBH at GC in the framework of Yukawa gravity. From the obtained values of $\Lambda$, and assuming that it corresponds to the Compton wavelength of graviton, we then calculated new estimates for the upper bound of graviton mass which are found to be independent, but consistent with the LIGO's estimate of graviton mass from the first gravitational wave (GW) signal GW150914. We also perform additional analysis using $f_{SP}$ parameter in order to constrain the bounds on graviton mass in the case of a small deviation of the stellar orbits from the corresponding GR predictions, as it is expected from the future more precise observations, and show that in such a case our method could further improve the previous estimates for upper bounds on the graviton. 2) We show that the current GRAVITY estimate of $f_{SP}$ can improve our previous constraints on the upper bound of graviton mass in about 2-3 times, but at the same time, it results with high contribution to the relative error. We also analyse theoretically possible future more precision observations and see if it will confirm the GR prediction for the Schwarzschild precession with factor $f_{SP}$ more close to 1, and how it affects the upper bound of graviton mass. 3) The methods used in our previous papers \cite{zakh16a,zakh18a,jova23} are different than in this paper. Here we study the stellar orbits around Sgr A* in massive gravity using two different PPN equations of motion. In their recent paper, the GRAVITY Collaboration used a modified PPN equation of motion to parametrize the effect of the Schwarzschild metric by introducing an ad hoc factor $f_{SP}$, characterizing how relativistic the model is, in front of the first post-Newtonian correction of GR. Besides the extended PPN formalism which we used in our previous \cite{jova23} and which contains an additional Yukawa-like term, here we also use the above modified PPN formalism presented by the GRAVITY Collaboration to study the stellar orbits in massive gravity and to constrain its range of interaction $\Lambda$. Here we also compared these two approaches and showed that they gave similar results. Both models produce the same pericenter advances per orbit, but their functional forms will still differ. For more information about PPN approach and detailed discussion see book by C. M. Will and references therein \cite{will18b}. 

This paper is organized as follows: in Section 2 we presented the PPN equations of motion and other important expressions that we used for analysis of the stellar orbits around Sgr A* in Yukawa gravity. We performed our analysis for $\delta$ = 1 and obtained results for upper bound of graviton mass in case of different S-stars. These results and the corresponding discussion are presented in Section 3, while Section 4 is devoted to the concluding remarks.

\begin{figure*}[ht!]
\centering
\includegraphics[width=0.58\linewidth]{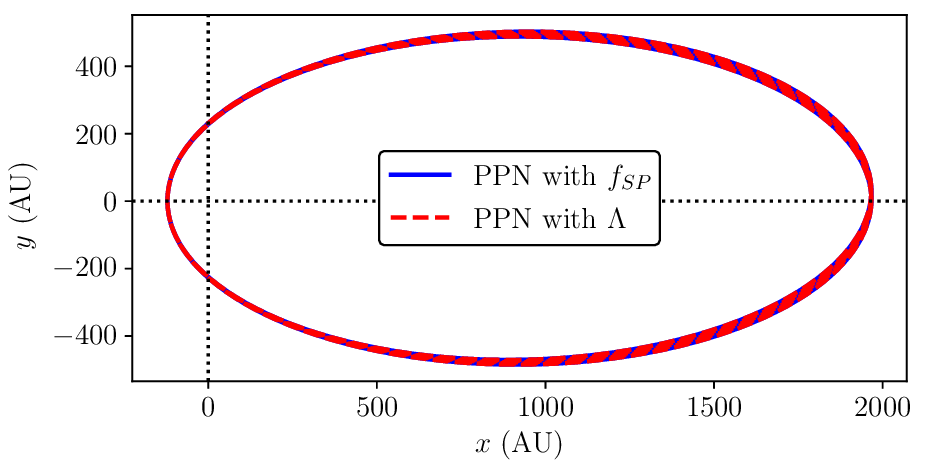}\hfill
\includegraphics[width=0.4\linewidth]{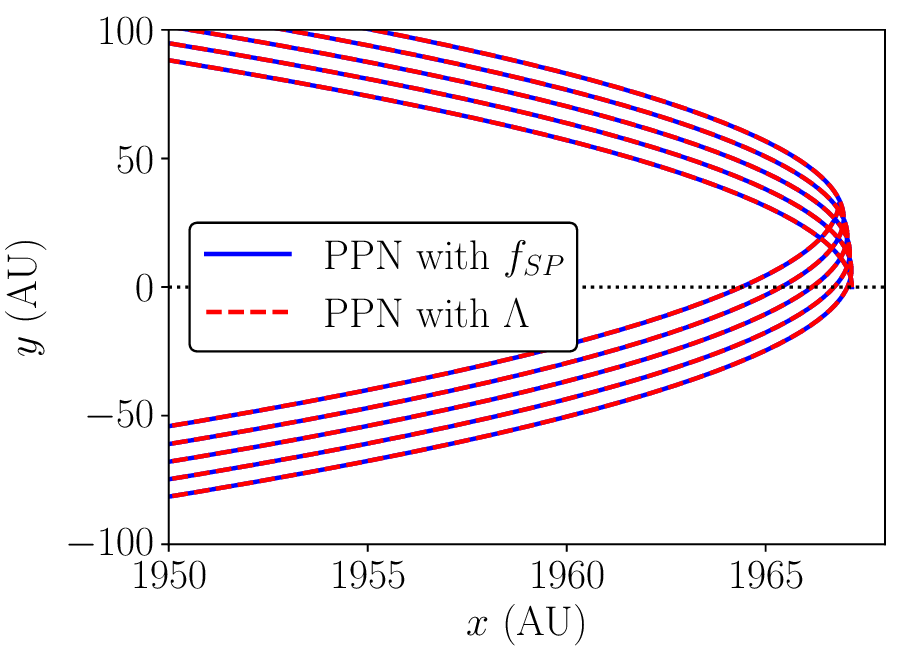}\\
\caption{Comparisons between the simulated orbits of S2 star, obtained by numerical integration of equation of motions in PPN formalism (\ref{ppneom1}) for $f_{SP}=1.10$ (blue solid line) and in PPN formalism (\ref{ppneom2}) for $\Lambda=46924.6$ AU (red dashed line). The orbits are calculated during five orbital periods, and their zoomed parts around the apocenter, where the largest discrepancy could occur, are presented in the right panel for better insight.}
\label{fig1}
\end{figure*}

\begin{figure*}[ht!]
	\centering
	\includegraphics[width=0.4\linewidth]{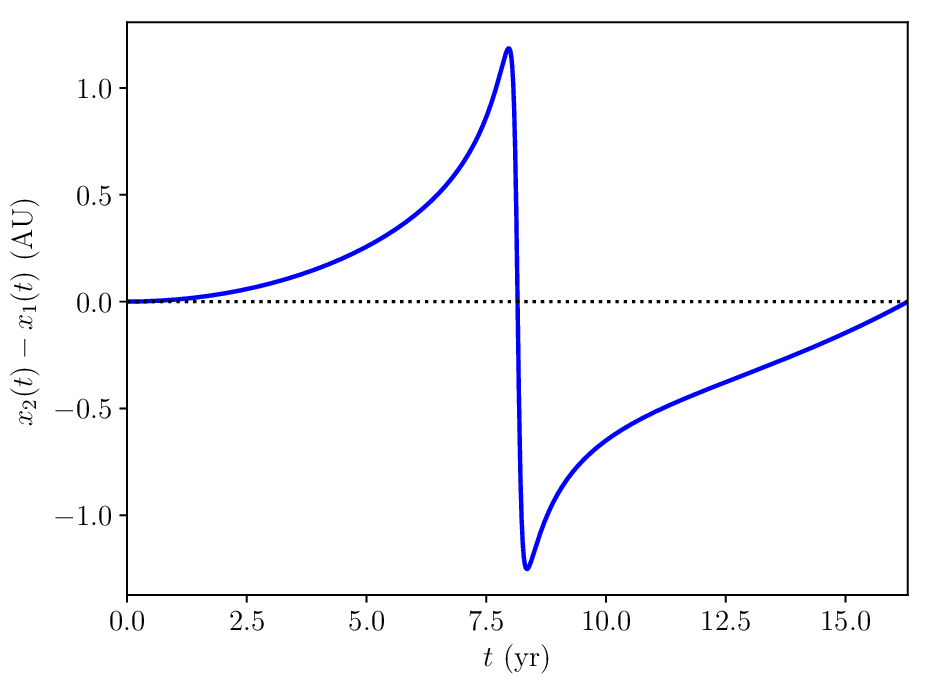}\hfill
	\includegraphics[width=0.4\linewidth]{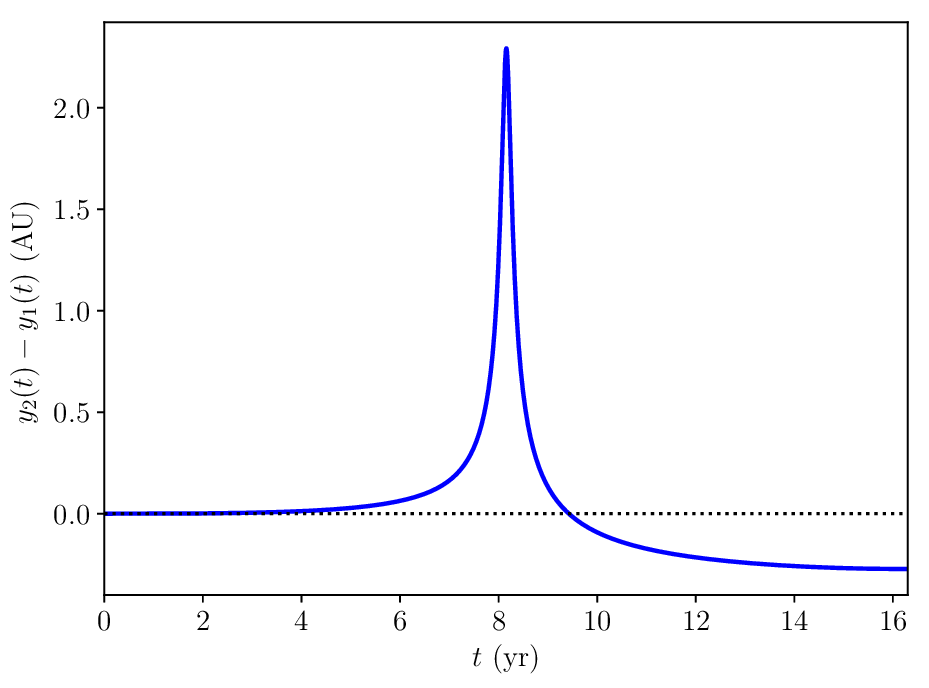}\\
	\caption{Time dependence of difference between simulated orbits accompanied to Fig. \ref{fig1} with panel (a): ($x(t)_2$ - $x(t)_1$) and panel (b): ($y(t)_2$ - $y(t)_1$), where the subscripts refers to the models (model 1 for Eq.(\ref{ppneom1}) for $f_{SP}=1.10$ and model 2 for Eq.(\ref{ppneom2}) for $\Lambda=46924.6$ AU, respectively, within one orbital period.}
	\label{fig2}
\end{figure*}

\begin{figure*}[ht!]
	\centering
	\includegraphics[width=0.4\linewidth]{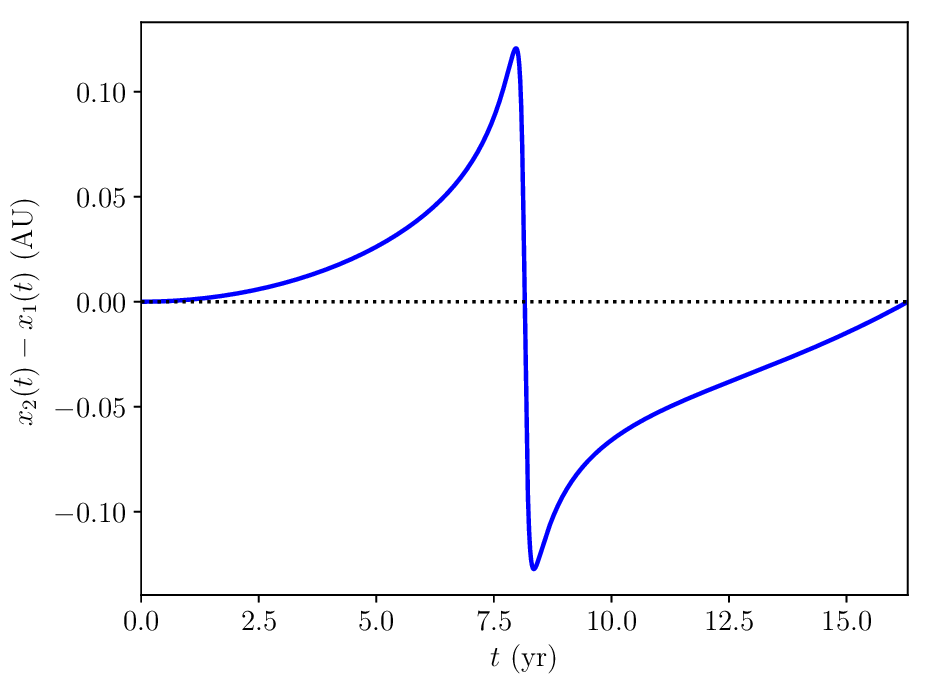}\hfill
	\includegraphics[width=0.4\linewidth]{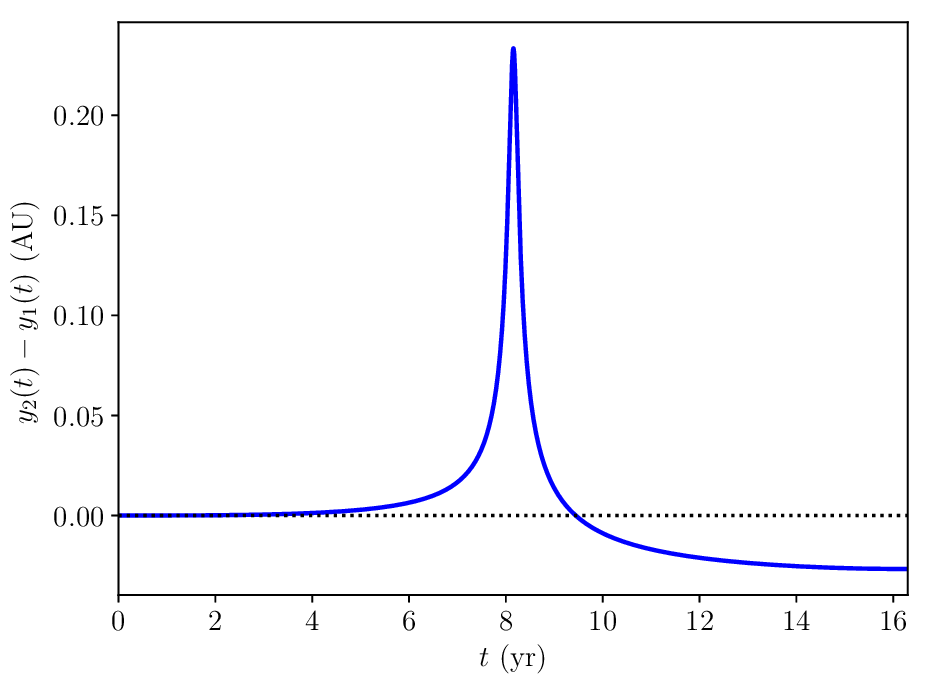}\\
	\caption{The same as in Fig. \ref{fig2} but for $f_{SP}=1.01$ and corresponding $\Lambda=148388.8$ AU}
	\label{fig3}
\end{figure*}

\begin{figure*}[ht!]
\centering
\includegraphics[width=0.58\linewidth]{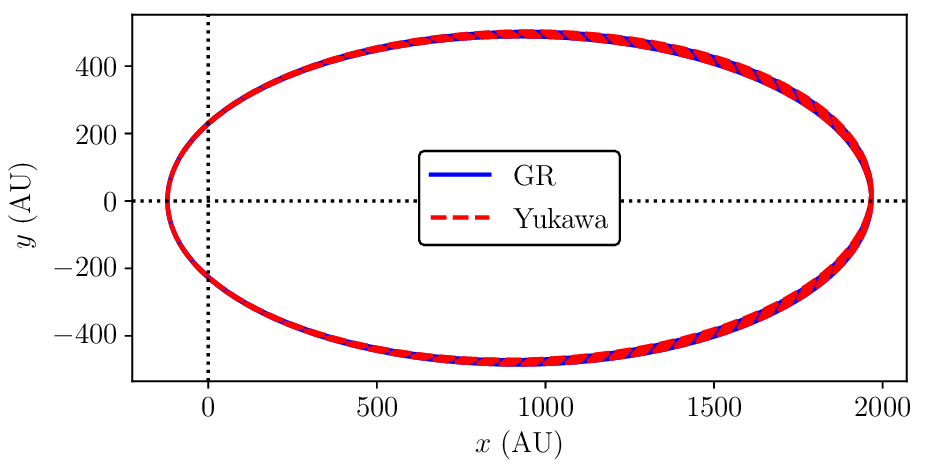}\hfill
\includegraphics[width=0.4\linewidth]{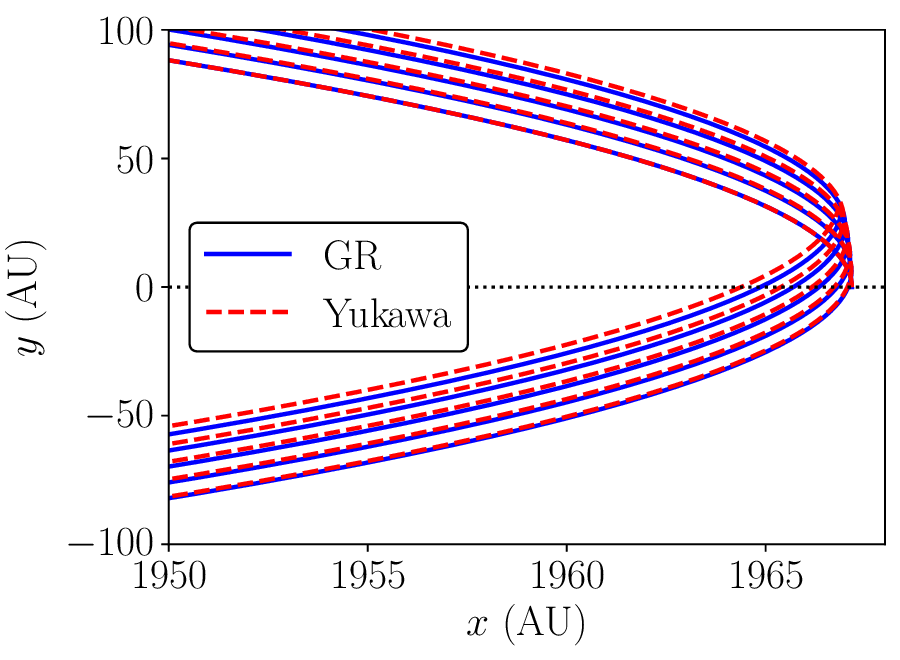}\\
\caption{Comparisons between the simulated orbits of S2 star, obtained by numerical integration of PPN equation in GR (blue solid line) and in Yukawa gravity for $\Lambda=46924.6$ AU (red dashed line, which corresponds to $f_{SP}=1.10$) using PPN formalism (\ref{ppneom2}). The orbits are calculated during five orbital periods, and their zoomed parts around the apocenter, where the largest discrepancy could occur, are presented in the right panel for better insight.}
\label{fig4}
\end{figure*}

\begin{figure*}[ht!]
\centering
\includegraphics[width=0.58\linewidth]{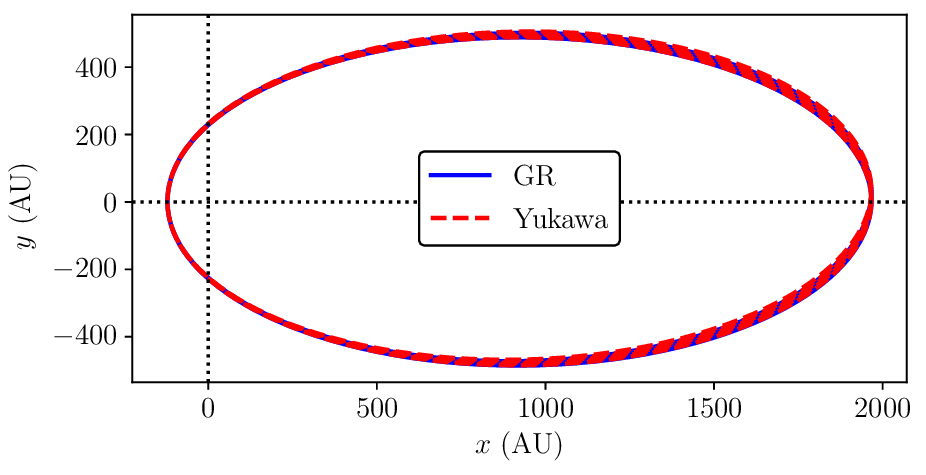}\hfill
\includegraphics[width=0.4\linewidth]{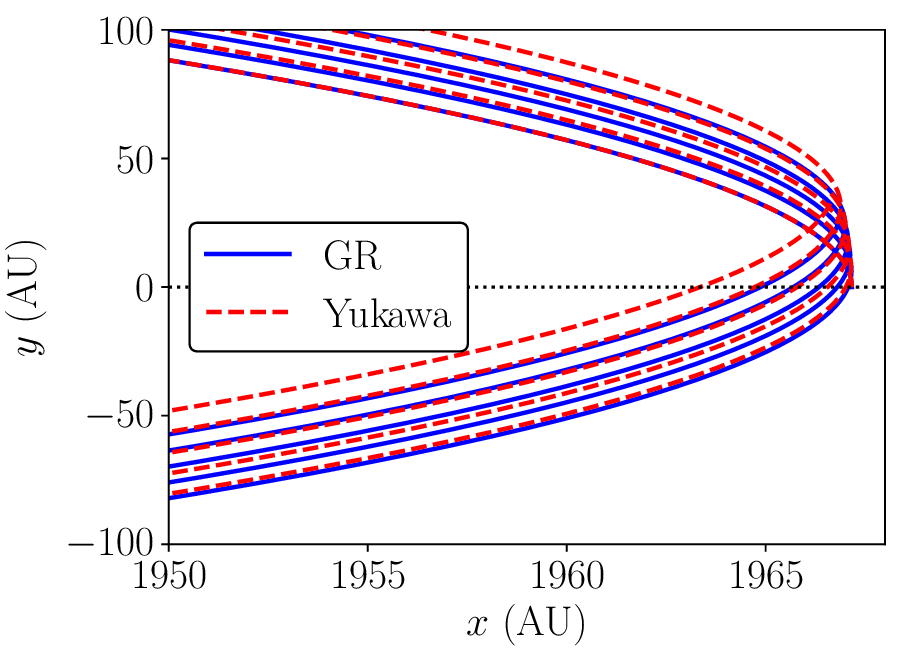}\\
\caption{The same as in Fig. \ref{fig2}, but for $\Lambda=27555.1$ AU which corresponds to $f_{SP}=1.29$.}
\label{fig5}
\end{figure*}

\section{Analysis of the stellar orbits around Sgr A* in massive gravity}

Here we study the stellar orbits around Sgr A* in massive gravity using two different PPN equations of motion. Namely, in recent paper \cite{abut20} the GRAVITY Collaboration used a modified PPN equation of motion to parametrize the effect of the Schwarzschild metric by introducing an ad hoc factor $f_{SP}$ in front of the first post-Newtonian correction of GR. This parameter is defined as $f_{SP} = (2 + 2\gamma - \beta) / 3$ , where $\beta$ and $\gamma$ are the post-Newtonian parameters which in the case of GR are both equal to 1, and thus $f_{SP}=1$ in this case. Therefore, $f_{SP}$ shows to which extent some gravitational model is relativistic. The value of $f_{SP} = 1.10 \pm 0.19$ was obtained by the GRAVITY Collaboration in the case of orbit of S2 star around Sgr A* \cite{abut20}. In that way they obtained the following modified PPN equation of motion:
\begin{equation}
\vec{\ddot{r}}=-GM \dfrac{\vec{r}}{r^3}+f_{SP}\dfrac{GM}{c^2r^3} \left[\left(4\dfrac{G M}{r}-\vec{\dot{r}}\cdot\vec{\dot{r}}\right) \vec{r} + 4\left(\vec{r}\cdot\vec{\dot{r}}\right)\vec{\dot{r}}\right].
\label{ppneom1}
\end{equation}
For $f_{SP}=1$ this expression reduces to the standard PPN equation of motion in GR.

On the other hand, in our recent paper, we used the so called extended PPN formalism for equation of motion in Yukawa gravity (see expression (3.9) in \cite{jova23}). This formalism contains an extra term which arises because the standard PPN formalism is not viable when dealing with theories that contain massive fields, and therefore requires a modification of the Newtonian order terms by a Yukawa-like correction (see e.g. \cite{clif08,alsi12}). The $f(R)$ gravity in the low energy limit gives the Yukawa potential (\ref{yukawa}), but also $f(R)$ gravity includes the first post-Newtonian approximations that in the limit $f(R)\rightarrow R$ should coincide with GR. Therefore, we consider a model that includes Yukawa correction + the post-Newtonian correction. Besides, the potential of the form $\Phi(r) = \dfrac{GM}{r}e^{-r/\lambda_g}$ which does not include $\delta$ is usually used in the frame of a massive graviton theory (see e.g. \cite{gold10,rana18,desa18,gupt18,cliff06,clif08,will98}). As noted in \cite{will98}, such a potential is a phenomenological assumption, and its exact form with Yukawa correction term should be obtained in the frame of some specific gravitational theory of a massive graviton. In the case of massive gravity theory derived from $f(R)$ theories (see e.g. Eqs. (2.1)-(2.7) in \cite{jova23}), such potential has the form given by Eq. (1). In order to obtain a form of this potential which does not depend of $\delta$, and taking into account that the goal of the present paper is to study the constraints on the graviton mass $m_g$, we assumed that $\delta$ = 1, which is a common assumption in such a case \cite{gold10,rana18,desa18,gupt18,cliff06,clif08,will98}. Thus, the PPN equation of motion in Yukawa gravity has the following form:
\begin{widetext}
\begin{equation}
\vec{\ddot{r}}=-GM \dfrac{\vec{r}}{r^3}+\dfrac{GM}{c^2r^3} \left[\left(4\dfrac{G M}{r}-\vec{\dot{r}}\cdot\vec{\dot{r}}\right) \vec{r} + 4\left(\vec{r}\cdot\vec{\dot{r}}\right)\vec{\dot{r}}\right]+\dfrac{GM}{2} \left[ 1 - \left(1 + \dfrac{r}{\Lambda} \right) e^{-\dfrac{r}{\Lambda}} \right] \dfrac{\vec{r}}{r^3}.
\label{ppneom2}
\end{equation}
\end{widetext}
The last term in r.h.s. of (\ref{ppneom2}) with Yukawa-like correction becomes negligible when $\Lambda\rightarrow\infty$, and then (\ref{ppneom2}) also reduces to the standard PPN equation of motion in GR.

Moreover, in the mentioned study \cite{jova23}, we also derived the following relation between the $\Lambda$ and $f_{SP}$:
\begin{equation}
\Lambda(P,e;f_{SP}) \approx \dfrac{cP}{4\pi} \sqrt{\dfrac{(\sqrt{1-e^2})^3}{3(f_{SP} - 1)}},
\label{lambda}
\end{equation}
for which the both PPN equations of motion (\ref{ppneom1}) and (\ref{ppneom2}) will result with practically the same orbital precession in the simulated orbits of S-stars. In the present paper we will exploit this fact to try to improve our previous constraints on the range of Yukawa gravity $\Lambda$ and graviton mass $m_g$ using the measured value of $f_{SP}$ obtained by the GRAVITY Collaboration. We chose the indirect path via Eq. (4) to constrain $\Lambda$ because the observational data which constrained $f_{SP}$ is not publicly available. 

Our method is based on (\ref{lambda}) and it represented the secular aspect of the evolution, in which the effects of both the models of Eqs. (\ref{ppneom1}) and (\ref{ppneom2}) are the pericentre advances $\Delta \omega$ per orbit. The method however neglects the non-secular aspects of the evolution, i.e., the continuous changes of the orbit as a function of time. That is why even in case that both models produce the same $\Delta\omega$ per orbit, their functional forms $\omega(t)$ will still differ (see Figs. \ref{fig1}, \ref{fig2} and \ref{fig3}). More discussion about non-secular effects one can find in recent papers \cite{heib22,alus22} and references therein. 

The orbital shift can be due to relativistic effects, resulting in a prograde shift, and due to a possible extended mass distribution, producing a retrograde shift. The retrograde Newtonian shift may partially or completely compensate the relativistic shift \cite{rubi01}. In case of S2 star, it was shown that the Schwarzschild precession dominates the entire orbit and that there is no detectable retrograde (Newtonian) precession due to an extended mass component (see \cite{heib22}). In general, the impact of an extended mass is naturally largest near apocenter of an orbit \cite{heib22}.

The relative error of Yukawa gravity parameter $\Lambda$ (and also of graviton mass $m_g$) can be obtained by differentiating the logarithmic versions of the above expression (\ref{lambda}):
\begin{equation}
\dfrac{|\Delta\Lambda|}{\Lambda}=\dfrac{|\Delta m_g|}{m_g} \leq \left(\dfrac{\left|\Delta P\right|}{P}+\dfrac{3 e\left|\Delta e\right|}{2 (1-e^2)}+\dfrac{\left|\Delta f_{SP}\right|}{2 (f_{SP}-1)}\right).
\label{relerr}
\end{equation}

Besides, assuming that the the range of Yukawa interaction $\Lambda$ corresponds to the graviton Compton wavelength $\lambda_g$:
\begin{equation}
\lambda_g=\dfrac{h c}{m_g},
\label{compton}
\end{equation}
the expression (\ref{lambda}) can be recast to obtain the orbital period $P$ as a function of eccentricity $e$ and two free parameters $m_g$ and $f_{SP}$:
\begin{equation}
P(e;m_g,f_{SP}) \approx \dfrac{4\pi h}{m_g} \sqrt{\dfrac{3(f_{SP} - 1)}{(\sqrt{1-e^2})^3}}.
\label{period}
\end{equation}

The expression (\ref{period}) could be easily recast in order to obtain the following dependence of graviton mass $m_g$ on the orbital periods $P$ and eccentricities $e$ of S-stars, for a specified value of the parameter $f_{SP}$:
\begin{equation}
m_g(P,e;f_{SP}) \approx \dfrac{4\pi h}{P} \sqrt{\dfrac{3(f_{SP} - 1)}{(\sqrt{1-e^2})^3}}.
\label{gravmass}
\end{equation}
This expression will enable us to obtain the new estimates for graviton mass $m_g$ which correspond to the given values of $f_{SP}$, using the observed orbital periods $P$ and eccentricities $e$ of S-stars. We will then compare these results with our previous constraints on graviton mass $m_g$ and study if there are any improvements with respect to our previous constraints from Table 2 in \cite{zakh18a}.

\begin{figure*}[ht!]
\centering
\includegraphics[width=0.52\linewidth]{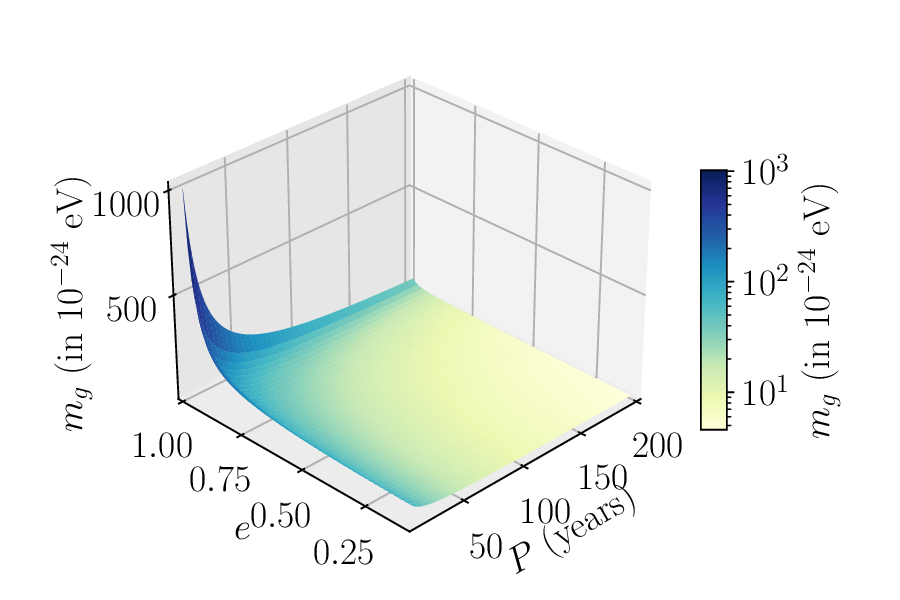}\hfill
\includegraphics[width=0.43\linewidth]{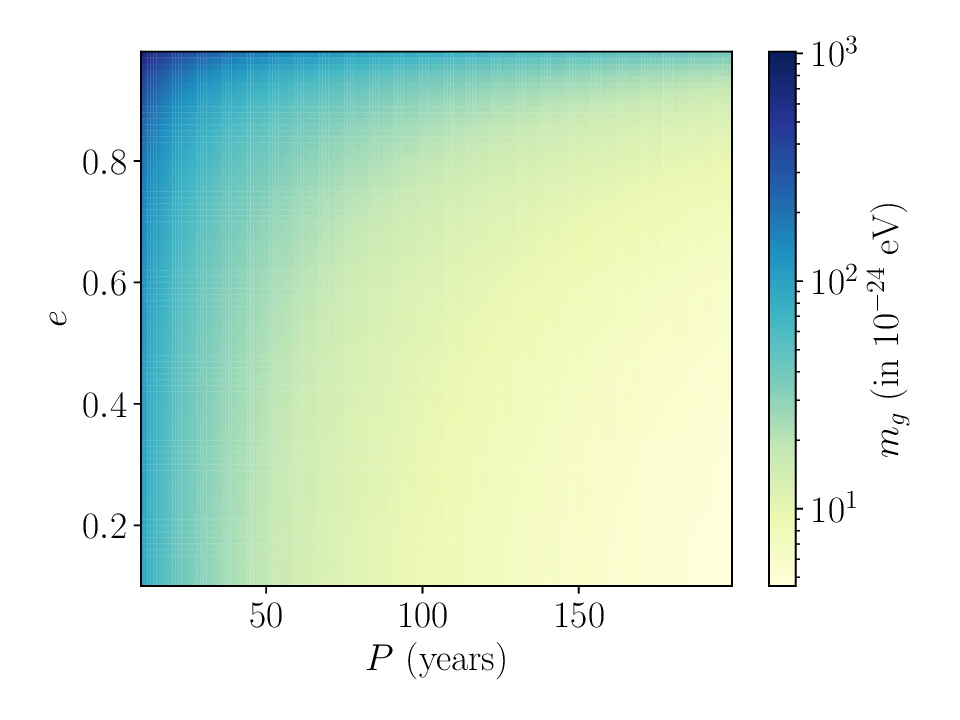}\\
\caption{The 3D graphic (left) of the function $m_g(P,e;f_{SP})$ given by (\ref{gravmass}), representing the dependence of graviton mass on the orbital periods $P$ and eccentricities $e$ of S-stars in the case of $f_{SP} = 1.10$, as well as the corresponding projection to the $P-e$ parameter space (right).}
\label{fig6}
\end{figure*}

\begin{figure*}[ht!]
\centering
\includegraphics[width=0.52\linewidth]{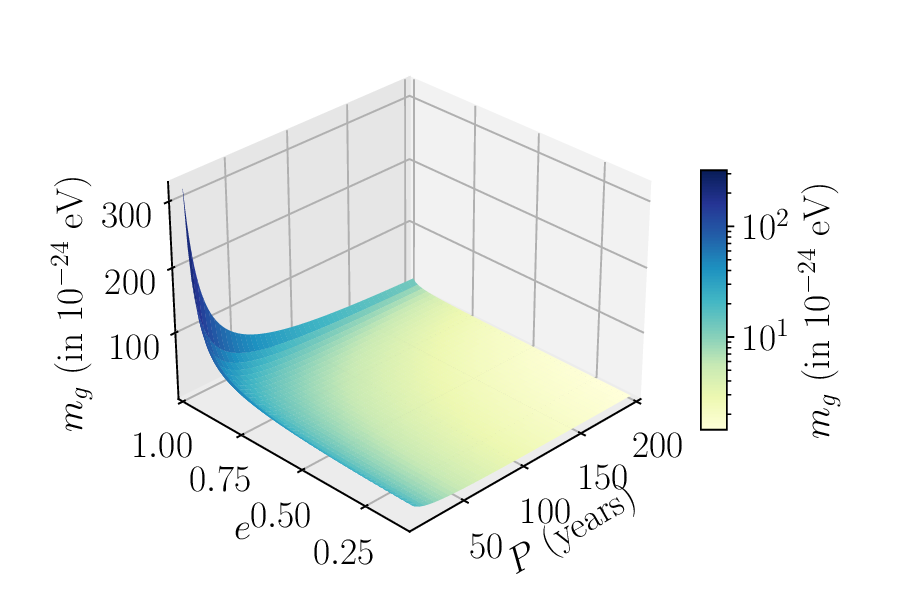}\hfill
\includegraphics[width=0.43\linewidth]{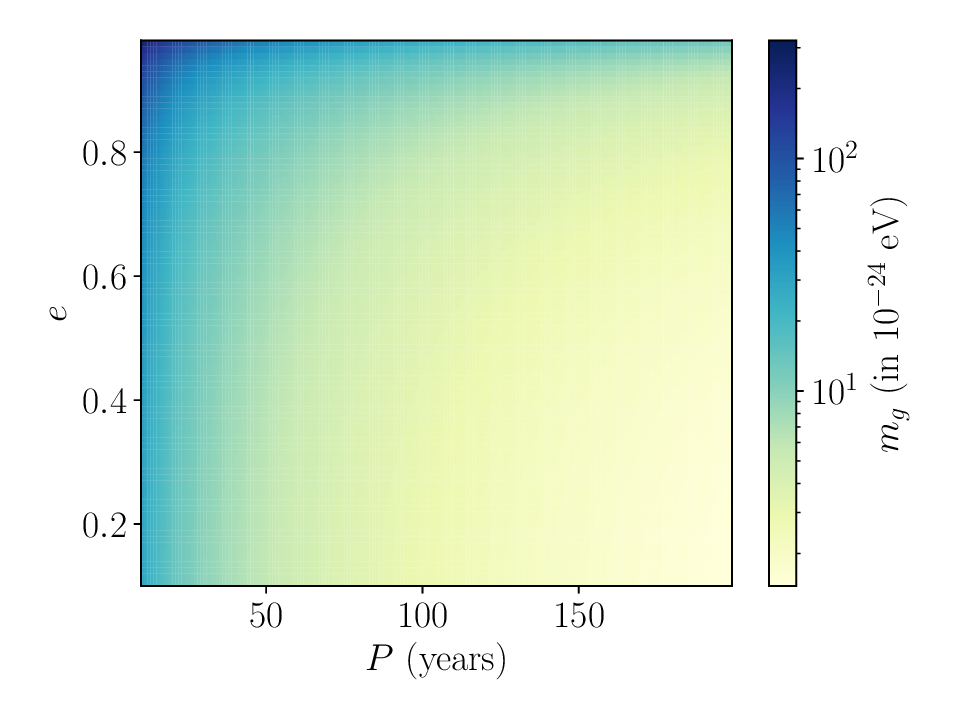}\\
\caption{The same as in Fig. \ref{fig6}, but for $f_{SP} = 1.01$.}
\label{fig7}
\end{figure*}

\begin{figure*}[ht!]
\centering
\includegraphics[width=0.52\linewidth]{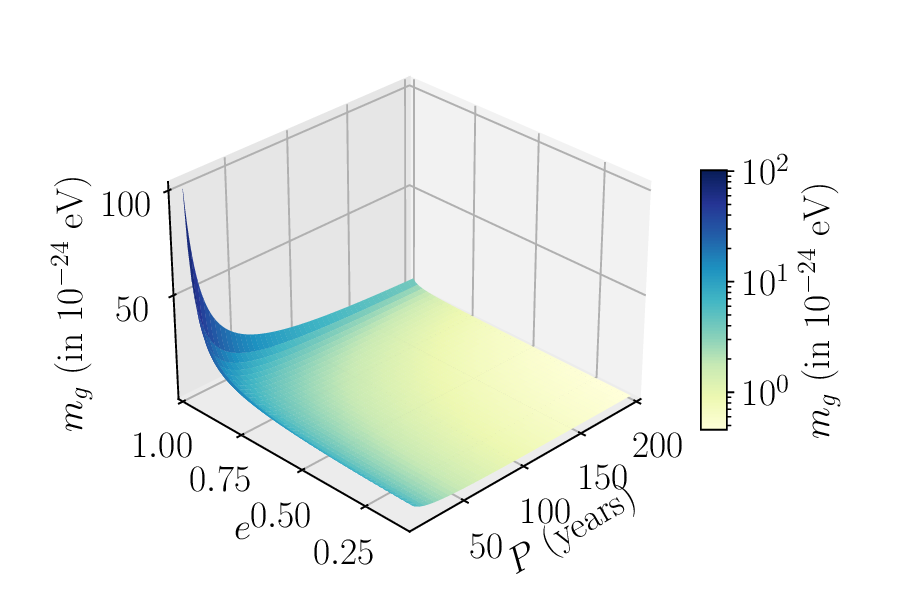}\hfill
\includegraphics[width=0.43\linewidth]{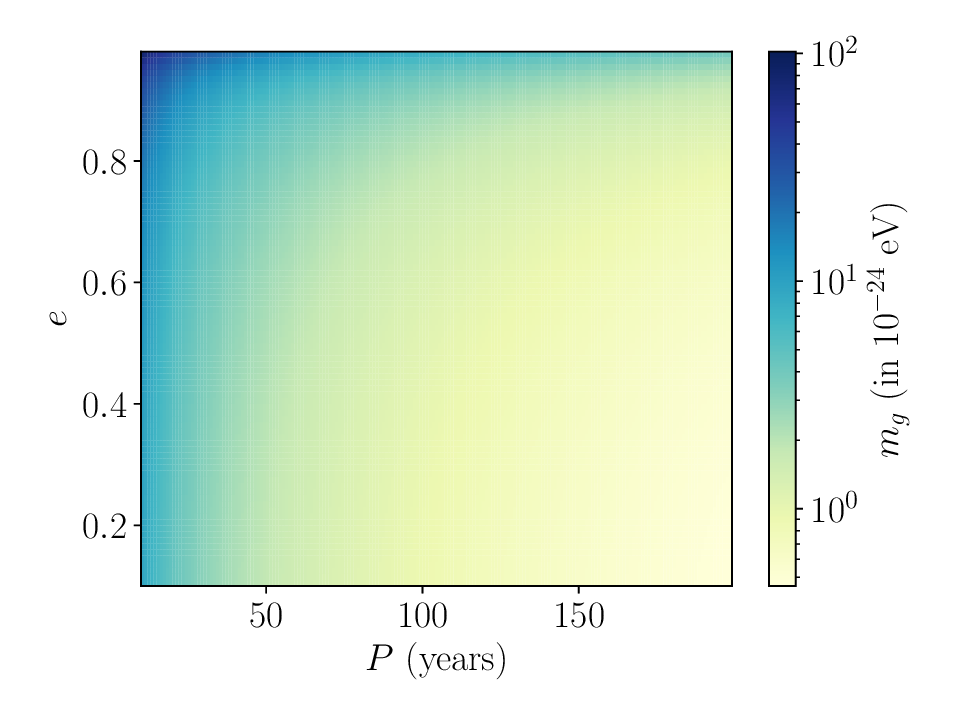}\\
\caption{The same as in Fig. \ref{fig6}, but for $f_{SP} = 1.001$.}
\label{fig8}
\end{figure*}

\section{Results and discussion}

In order to test whether the condition (\ref{lambda}) will also result with the sufficiently close simulated orbits of S-stars in the both studied PPN formalisms, we calculated the simulated orbits of S2 star by numerical integration of equation of motions Eq.(\ref{ppneom1}) and Eq.(\ref{ppneom2}) for $f_{SP} = 1.10$ and $\Lambda = 46924.6$ AU that correspond to each other in the case of S2 star. The value $\Lambda = 46924.6$ AU is obtained by Eq.(\ref{lambda}) when we put $f_{SP} = 1.10$ and take values of the period and eccentricity of S2 star, P=16.00 yr and e=0.8839, respectively. Both values are taken from Table 3 in \cite{gill17}, which presented observed data with corresponding errors. Both orbits are calculated by numerical integration of the corresponding equation of motion, using the Keplerian positions and velocities at apocenter as initial conditions. The obtained simulated orbits of S2 star are presented in Fig. \ref{fig1}, from which it can be seen that these orbits practically overlap. In this way, we numerically tested the validity of Eq. (4) and demonstrated that it holds in the case of the two-body problem.
In order to provide information about time dependence we accompany Fig. \ref{fig1} with new plots ($x(t)_2$ - $x(t)_1$) and ($y(t)_2$ - $y(t)_1$), presented in Fig. \ref{fig2}, where the subscripts refers to the models (model 1 for Eq.(\ref{ppneom1}) for $f_{SP}=1.10$ and model 2 for Eq.(\ref{ppneom2}) for $\Lambda=46924.6$ AU) within one orbital period. From Figs. \ref{fig1} and \ref{fig2} we can conclude that obtained simulated orbits practically overlap, but a small difference exists. Fig. \ref{fig3} presented time dependence of difference between simulated orbits in the model 1 for Eq.(\ref{ppneom1}) for $f_{SP}=1.01$ and model 2 for Eq.(\ref{ppneom2}) for the corresponding $\Lambda=148388.8$ AU, within one orbital period. From Figs. \ref{fig2} and \ref{fig3} we can conclude that obtained simulated orbits in model 1 and model 2 have smaller difference between each other when $f_{SP}$ is closer to the GR case.

Furthermore, we also compared the simulated orbits of S2 star obtained using the standard PPN equation of motion in GR (PPN formalism Eq.(\ref{ppneom1})) for $f_{SP} = 1$) with the corresponding orbits in Yukawa gravity obtained using the PPN formalism Eq.(\ref{ppneom2}) for $\Lambda = 46924.6$ AU which corresponds to $f_{SP}=1.10$, as well as for $\Lambda = 27555.1$ AU which corresponds to $f_{SP}=1.29$, i.e. to the upper limit within the error interval obtained by the GRAVITY collaboration: $f_{sp} = 1.10 + 0.19 = 1.29$ \cite{abut20}. These results are presented in Figs. \ref{fig4} and \ref{fig5}, respectively, and it can be seen that the difference between the orbits in Yukawa gravity from those in GR becomes more noticeable with decrease of $\Lambda$ corresponding to the increasing deviations of $f_{SP}$ from 1. All this indicates that the condition (\ref{lambda}) could be used for improving the constraints on the range of Yukawa gravity $\Lambda$ and graviton mass $m_g$ using the latest estimates for $f_{SP}$.

We used the Yukawa gravity as a tool to estimate the current graviton mass bounds, as well as those under a scenario in which GR prediction for $f_{SP}$ will be confirmed with much higher accuracy by future more precise observations. When compared with our previous estimate of the upper bound of graviton mass $m_g < 4\times 10^{-23}$ eV \cite{zakh18a}, it can be seen that if the future high precision observations will confirm GR prediction for Schwarzschild precession with factor $f_{SP}$, it could significantly improve these constraints, achieving even $\sim 6$ times better estimates in case $f_{SP}=1.01$, or for closer values of $f_{SP}$ to 1, i.e. $f_{SP}=1.001$, even $\sim 15$ times.

In order to study the dependence of graviton mass $m_g$ on the orbital periods $P$ and eccentricities $e$ of S-stars, we plotted the function $m_g(P,e;f_{SP})$ given by Eq. (\ref{gravmass}) in the form of a 3D graphic over the $P-e$ parameter space. Three such 3D graphics of this function in the cases of $f_{SP}$ = 1.10, 1.01 and 1.001 are presented in the left panels of the Figs. \ref{fig6} - \ref{fig8}, respectively, and the corresponding projections to the $P-e$ parameter space are given in the right panels of these figures. As it can be seen from Figs. \ref{fig6} - \ref{fig8}, the smaller values of the graviton mass $m_g$ are obtained for the S-stars with larger orbital periods $P$ and lower eccentricities $e$, which is also in agreement with our recent results for the range of Yukawa interaction $\Lambda$ (see Fig. 1 in \cite{jova23}), but at the same time the resulting precession becomes smaller and thus harder to observe. Also, larger value of period $P$ requires longer monitoring, sometimes for a few hundred years (see data for $P$ from from Table 3 in \cite{gill17}). The second important issue is that above claim is correct only under the assumption (which is made here) that $f_{SP}$ has been measured for such an orbit already to a given precision, but in real situation it is expected that different orbits have slightly different accuracies of measurement of $f_{SP}$. Different orbits are differently suited to constrain$f_{SP}$, and the one of S2 happens to be particularly good for that, i.e. this orbit has (high $e$, low $P$) and it is successfully measured by GRAVITY collaboration. Our assumption that the same accuracy of measurement of $f_{SP}$ for all the S-stars in Tables \ref{tab1} in reality probably does not hold for all S-stars. Probably, differences in accuracy of measurements of $f_{SP}$ are less for orbits if they have relatively small mutual discrepancies in $e$ and $P$, so S-stars with orbits similar to S2 orbit (high $e$, short $P$) will have $f_{SP}$ measured with very similar precision.

Besides, the shape of the surface representing the function $m_g(P,e;f_{SP})$ is similar for all studied values of $f_{SP}$, but the values of $m_g$ are $\sim 3$ or $\sim 10$ times of magnitude smaller for $f_{SP}$ = 1.01 or 1.001 with respect of those in the case of $f_{SP} = 1.10$ (see Figs. \ref{fig6} - \ref{fig8} and Tables \ref{tab1} and \ref{tab2}). In the latter case, graviton mass could go well bellow $1\times 10^{-24}$ eV (see Table \ref{tab2}). This confirms our previous result that the future high precision observations could significantly improve the existing constraints on the upper bound of graviton mass, if they manage to confirm the GR prediction for Schwarzschild precession with $f_{SP} = 1.01$, or much smaller.

%\vfil\clearpage

\begin{table*}
\centering
\caption{The range of Yukawa interaction $\Lambda$, graviton mass $m_g$, as well as their relative and absolute errors, calculated for $f_{SP} = 1.10\pm0.19$ and for $f_{SP} = 1.19\pm0.19$ in the case of all S-stars from Table 3 in \cite{gill17} except of S111.}
\label{tab1}
	\setlength{\tabcolsep}{0.11cm}
	\begin{tabular}{|l|rcl|rcl|r|rcl|rcl|r|}
		\hline
		&\multicolumn{7}{c|}{$f_{SP}=1.10\pm0.19$}&\multicolumn{7}{c|}{$f_{SP}=1.19\pm0.19$}\\
		\cline{2-15}
		\multicolumn{1}{|l|}{Star}&\multicolumn{3}{c|}{$\Lambda\pm\Delta\Lambda$}&\multicolumn{3}{c|}{$m_g\pm\Delta m_g$}&\multicolumn{1}{c|}{R.E.}&\multicolumn{3}{c|}{$\Lambda\pm\Delta\Lambda$}&\multicolumn{3}{c|}{$m_g\pm\Delta m_g$}&\multicolumn{1}{c|}{R.E.} \\
		&\multicolumn{3}{c|}{(AU)}&\multicolumn{3}{c|}{$(10^{-24}\ \mathrm{eV})$}&\multicolumn{1}{c|}{(\%)}&\multicolumn{3}{c|}{(AU)}&\multicolumn{3}{c|}{$(10^{-24}\ \mathrm{eV})$}&\multicolumn{1}{c|}{(\%)} \\
		\hline
		\hline
		S1   &   1.2e+06 & $\pm$ & 1.2e+06 &   7.2 & $\pm$ & 7.2 & 100.7 &   8.4e+05 & $\pm$ & 4.7e+05 &   9.9 & $\pm$ & 5.5 &  55.7 \\
		S2   &   4.7e+04 & $\pm$ & 4.5e+04 & 176.6 & $\pm$ & 170.0 &  96.3 &   3.4e+04 & $\pm$ & 1.7e+04 & 243.5 & $\pm$ & 124.8 &  51.3 \\
		S4   &   6.2e+05 & $\pm$ & 6.0e+05 &  13.3 & $\pm$ & 12.9 &  96.7 &   4.5e+05 & $\pm$ & 2.3e+05 &  18.3 & $\pm$ & 9.5 &  51.7 \\
		S6   &   7.0e+05 & $\pm$ & 6.7e+05 &  11.8 & $\pm$ & 11.2 &  95.2 &   5.1e+05 & $\pm$ & 2.6e+05 &  16.2 & $\pm$ & 8.2 &  50.2 \\
		S8   &   3.9e+05 & $\pm$ & 3.8e+05 &  21.2 & $\pm$ & 20.7 &  98.0 &   2.8e+05 & $\pm$ & 1.5e+05 &  29.2 & $\pm$ & 15.4 &  53.0 \\
		S9   &   3.1e+05 & $\pm$ & 3.1e+05 &  26.3 & $\pm$ & 26.2 &  99.7 &   2.3e+05 & $\pm$ & 1.2e+05 &  36.3 & $\pm$ & 19.8 &  54.7 \\
		S12  &   1.7e+05 & $\pm$ & 1.6e+05 &  49.3 & $\pm$ & 47.6 &  96.4 &   1.2e+05 & $\pm$ & 6.3e+04 &  68.0 & $\pm$ & 35.0 &  51.4 \\
		S13  &   3.9e+05 & $\pm$ & 3.7e+05 &  21.4 & $\pm$ & 20.4 &  95.5 &   2.8e+05 & $\pm$ & 1.4e+05 &  29.5 & $\pm$ & 14.9 &  50.5 \\
		S14  &   5.1e+04 & $\pm$ & 5.5e+04 & 161.3 & $\pm$ & 173.2 & 107.3 &   3.7e+04 & $\pm$ & 2.3e+04 & 222.4 & $\pm$ & 138.6 &  62.3 \\
		S17  &   6.2e+05 & $\pm$ & 6.0e+05 &  13.4 & $\pm$ & 13.0 &  97.1 &   4.5e+05 & $\pm$ & 2.3e+05 &  18.5 & $\pm$ & 9.6 &  52.1 \\
		S18  &   3.2e+05 & $\pm$ & 3.1e+05 &  26.0 & $\pm$ & 25.1 &  96.5 &   2.3e+05 & $\pm$ & 1.2e+05 &  35.9 & $\pm$ & 18.5 &  51.5 \\
		S19  &   6.7e+05 & $\pm$ & 7.8e+05 &  12.4 & $\pm$ & 14.5 & 116.4 &   4.8e+05 & $\pm$ & 3.5e+05 &  17.2 & $\pm$ & 12.3 &  71.4 \\
		S21  &   1.8e+05 & $\pm$ & 1.8e+05 &  47.1 & $\pm$ & 46.9 &  99.6 &   1.3e+05 & $\pm$ & 7.0e+04 &  65.0 & $\pm$ & 35.5 &  54.6 \\
		S22  &   4.2e+06 & $\pm$ & 4.8e+06 &   2.0 & $\pm$ & 2.3 & 114.1 &   3.0e+06 & $\pm$ & 2.1e+06 &   2.7 & $\pm$ & 1.9 &  69.1 \\
		S23  &   3.2e+05 & $\pm$ & 3.7e+05 &  26.2 & $\pm$ & 30.3 & 115.6 &   2.3e+05 & $\pm$ & 1.6e+05 &  36.1 & $\pm$ & 25.5 &  70.6 \\
		S24  &   8.9e+05 & $\pm$ & 9.2e+05 &   9.3 & $\pm$ & 9.6 & 103.2 &   6.5e+05 & $\pm$ & 3.8e+05 &  12.8 & $\pm$ & 7.5 &  58.2 \\
		S29  &   5.3e+05 & $\pm$ & 5.7e+05 &  15.8 & $\pm$ & 17.2 & 109.1 &   3.8e+05 & $\pm$ & 2.4e+05 &  21.7 & $\pm$ & 13.9 &  64.1 \\
		S31  &   7.6e+05 & $\pm$ & 7.3e+05 &  11.0 & $\pm$ & 10.6 &  96.4 &   5.5e+05 & $\pm$ & 2.8e+05 &  15.1 & $\pm$ & 7.8 &  51.4 \\
		S33  &   1.2e+06 & $\pm$ & 1.3e+06 &   6.7 & $\pm$ & 7.1 & 107.0 &   9.0e+05 & $\pm$ & 5.6e+05 &   9.2 & $\pm$ & 5.7 &  62.0 \\
		S38  &   7.6e+04 & $\pm$ & 7.3e+04 & 108.8 & $\pm$ & 103.7 &  95.4 &   5.5e+04 & $\pm$ & 2.8e+04 & 149.9 & $\pm$ & 75.5 &  50.4 \\
		S39  &   1.8e+05 & $\pm$ & 1.7e+05 &  47.0 & $\pm$ & 46.4 &  98.8 &   1.3e+05 & $\pm$ & 6.9e+04 &  64.7 & $\pm$ & 34.8 &  53.8 \\
		S42  &   2.3e+06 & $\pm$ & 2.8e+06 &   3.6 & $\pm$ & 4.4 & 122.7 &   1.7e+06 & $\pm$ & 1.3e+06 &   5.0 & $\pm$ & 3.9 &  77.7 \\
		S54  &   1.3e+06 & $\pm$ & 2.5e+06 &   6.3 & $\pm$ & 11.8 & 188.3 &   9.6e+05 & $\pm$ & 1.4e+06 &   8.7 & $\pm$ & 12.4 & 143.3 \\
		S55  &   6.8e+04 & $\pm$ & 6.6e+04 & 122.4 & $\pm$ & 119.4 &  97.6 &   4.9e+04 & $\pm$ & 2.6e+04 & 168.7 & $\pm$ & 88.7 &  52.6 \\
		S60  &   4.6e+05 & $\pm$ & 4.5e+05 &  17.9 & $\pm$ & 17.5 &  97.7 &   3.4e+05 & $\pm$ & 1.8e+05 &  24.6 & $\pm$ & 13.0 &  52.7 \\
		S66  &   6.0e+06 & $\pm$ & 6.1e+06 &   1.4 & $\pm$ & 1.4 & 101.4 &   4.4e+06 & $\pm$ & 2.5e+06 &   1.9 & $\pm$ & 1.1 &  56.4 \\
		S67  &   3.7e+06 & $\pm$ & 3.7e+06 &   2.2 & $\pm$ & 2.2 & 100.1 &   2.7e+06 & $\pm$ & 1.5e+06 &   3.1 & $\pm$ & 1.7 &  55.1 \\
		S71  &   9.2e+05 & $\pm$ & 9.9e+05 &   9.0 & $\pm$ & 9.7 & 107.3 &   6.7e+05 & $\pm$ & 4.2e+05 &  12.4 & $\pm$ & 7.7 &  62.3 \\
		S83  &   5.4e+06 & $\pm$ & 6.0e+06 &   1.5 & $\pm$ & 1.7 & 110.3 &   3.9e+06 & $\pm$ & 2.6e+06 &   2.1 & $\pm$ & 1.4 &  65.3 \\
		S85  &   1.6e+07 & $\pm$ & 3.4e+07 &   0.5 & $\pm$ & 1.1 & 211.0 &   1.2e+07 & $\pm$ & 2.0e+07 &   0.7 & $\pm$ & 1.2 & 166.0 \\
		S87  &   1.4e+07 & $\pm$ & 1.5e+07 &   0.6 & $\pm$ & 0.6 & 102.4 &   1.0e+07 & $\pm$ & 6.0e+06 &   0.8 & $\pm$ & 0.5 &  57.4 \\
		S89  &   2.5e+06 & $\pm$ & 2.7e+06 &   3.3 & $\pm$ & 3.6 & 107.8 &   1.8e+06 & $\pm$ & 1.1e+06 &   4.5 & $\pm$ & 2.9 &  62.8 \\
		S91  &   8.2e+06 & $\pm$ & 8.3e+06 &   1.0 & $\pm$ & 1.0 & 101.9 &   5.9e+06 & $\pm$ & 3.4e+06 &   1.4 & $\pm$ & 0.8 &  56.9 \\
		S96  &   5.9e+06 & $\pm$ & 5.9e+06 &   1.4 & $\pm$ & 1.4 & 100.0 &   4.3e+06 & $\pm$ & 2.4e+06 &   1.9 & $\pm$ & 1.1 &  55.0 \\
		S97  &   1.1e+07 & $\pm$ & 1.3e+07 &   0.8 & $\pm$ & 1.0 & 125.9 &   7.7e+06 & $\pm$ & 6.2e+06 &   1.1 & $\pm$ & 0.9 &  80.9 \\
		S145 &   3.1e+06 & $\pm$ & 4.3e+06 &   2.6 & $\pm$ & 3.6 & 136.7 &   2.3e+06 & $\pm$ & 2.1e+06 &   3.6 & $\pm$ & 3.3 &  91.7 \\
		S175 &   5.8e+04 & $\pm$ & 6.4e+04 & 143.4 & $\pm$ & 158.1 & 110.3 &   4.2e+04 & $\pm$ & 2.7e+04 & 197.6 & $\pm$ & 129.0 &  65.3 \\
		R34  &   5.4e+06 & $\pm$ & 6.5e+06 &   1.5 & $\pm$ & 1.8 & 120.5 &   3.9e+06 & $\pm$ & 3.0e+06 &   2.1 & $\pm$ & 1.6 &  75.5 \\
		R44  &   2.4e+07 & $\pm$ & 3.7e+07 &   0.4 & $\pm$ & 0.5 & 156.2 &   1.7e+07 & $\pm$ & 1.9e+07 &   0.5 & $\pm$ & 0.5 & 111.2 \\
		\hline
	\end{tabular}
\end{table*}

\begin{table*}
\centering
\caption{The same as in Table \ref{tab1} but for $f_{SP}=1.01\pm0.005$ and $f_{SP} = 1.001\pm0.0005$.}
\label{tab2}
\setlength{\tabcolsep}{0.11cm}
\begin{tabular}{|l|rcl|rcl|r|rcl|rcl|r|}
\hline
&\multicolumn{7}{c|}{$f_{SP}=1.01\pm0.005$}&\multicolumn{7}{c|}{$f_{SP}=1.001\pm0.0005$}\\
\cline{2-15}
\multicolumn{1}{|l|}{Star}&\multicolumn{3}{c|}{$\Lambda\pm\Delta\Lambda$}&\multicolumn{3}{c|}{$m_g\pm\Delta m_g$}&\multicolumn{1}{c|}{R.E.}&\multicolumn{3}{c|}{$\Lambda\pm\Delta\Lambda$}&\multicolumn{3}{c|}{$m_g\pm\Delta m_g$}&\multicolumn{1}{c|}{R.E.} \\
&\multicolumn{3}{c|}{(AU)}&\multicolumn{3}{c|}{$(10^{-24}\ \mathrm{eV})$}&\multicolumn{1}{c|}{(\%)}&\multicolumn{3}{c|}{(AU)}&\multicolumn{3}{c|}{$(10^{-24}\ \mathrm{eV})$}&\multicolumn{1}{c|}{(\%)} \\
\hline
\hline
	S1   &   3.6e+06 & $\pm$ & 1.1e+06 &   2.3 & $\pm$ & 0.7 &  30.7 &   1.2e+07 & $\pm$ & 3.5e+06 &   0.7 & $\pm$ & 0.2 &  30.7 \\
	S2   &   1.5e+05 & $\pm$ & 3.9e+04 &  55.9 & $\pm$ & 14.7 &  26.3 &   4.7e+05 & $\pm$ & 1.2e+05 &  17.7 & $\pm$ & 4.6 &  26.3 \\
	S4   &   2.0e+06 & $\pm$ & 5.3e+05 &   4.2 & $\pm$ & 1.1 &  26.7 &   6.2e+06 & $\pm$ & 1.7e+06 &   1.3 & $\pm$ & 0.4 &  26.7 \\
	S6   &   2.2e+06 & $\pm$ & 5.6e+05 &   3.7 & $\pm$ & 0.9 &  25.2 &   7.0e+06 & $\pm$ & 1.8e+06 &   1.2 & $\pm$ & 0.3 &  25.2 \\
	S8   &   1.2e+06 & $\pm$ & 3.5e+05 &   6.7 & $\pm$ & 1.9 &  28.0 &   3.9e+06 & $\pm$ & 1.1e+06 &   2.1 & $\pm$ & 0.6 &  28.0 \\
	S9   &   1.0e+06 & $\pm$ & 3.0e+05 &   8.3 & $\pm$ & 2.5 &  29.7 &   3.1e+06 & $\pm$ & 9.3e+05 &   2.6 & $\pm$ & 0.8 &  29.7 \\
	S12  &   5.3e+05 & $\pm$ & 1.4e+05 &  15.6 & $\pm$ & 4.1 &  26.4 &   1.7e+06 & $\pm$ & 4.4e+05 &   4.9 & $\pm$ & 1.3 &  26.4 \\
	S13  &   1.2e+06 & $\pm$ & 3.1e+05 &   6.8 & $\pm$ & 1.7 &  25.5 &   3.9e+06 & $\pm$ & 9.9e+05 &   2.1 & $\pm$ & 0.5 &  25.5 \\
	S14  &   1.6e+05 & $\pm$ & 6.1e+04 &  51.0 & $\pm$ & 19.0 &  37.3 &   5.1e+05 & $\pm$ & 1.9e+05 &  16.1 & $\pm$ & 6.0 &  37.3 \\
	S17  &   2.0e+06 & $\pm$ & 5.3e+05 &   4.2 & $\pm$ & 1.1 &  27.1 &   6.2e+06 & $\pm$ & 1.7e+06 &   1.3 & $\pm$ & 0.4 &  27.1 \\
	S18  &   1.0e+06 & $\pm$ & 2.7e+05 &   8.2 & $\pm$ & 2.2 &  26.5 &   3.2e+06 & $\pm$ & 8.4e+05 &   2.6 & $\pm$ & 0.7 &  26.5 \\
	S19  &   2.1e+06 & $\pm$ & 9.8e+05 &   3.9 & $\pm$ & 1.8 &  46.4 &   6.7e+06 & $\pm$ & 3.1e+06 &   1.2 & $\pm$ & 0.6 &  46.4 \\
	S21  &   5.6e+05 & $\pm$ & 1.6e+05 &  14.9 & $\pm$ & 4.4 &  29.6 &   1.8e+06 & $\pm$ & 5.2e+05 &   4.7 & $\pm$ & 1.4 &  29.6 \\
	S22  &   1.3e+07 & $\pm$ & 5.8e+06 &   0.6 & $\pm$ & 0.3 &  44.1 &   4.2e+07 & $\pm$ & 1.8e+07 &   0.2 & $\pm$ & 0.1 &  44.1 \\
	S23  &   1.0e+06 & $\pm$ & 4.6e+05 &   8.3 & $\pm$ & 3.8 &  45.6 &   3.2e+06 & $\pm$ & 1.4e+06 &   2.6 & $\pm$ & 1.2 &  45.6 \\
	S24  &   2.8e+06 & $\pm$ & 9.4e+05 &   2.9 & $\pm$ & 1.0 &  33.2 &   8.9e+06 & $\pm$ & 3.0e+06 &   0.9 & $\pm$ & 0.3 &  33.2 \\
	S29  &   1.7e+06 & $\pm$ & 6.5e+05 &   5.0 & $\pm$ & 1.9 &  39.1 &   5.3e+06 & $\pm$ & 2.1e+06 &   1.6 & $\pm$ & 0.6 &  39.1 \\
	S31  &   2.4e+06 & $\pm$ & 6.3e+05 &   3.5 & $\pm$ & 0.9 &  26.4 &   7.6e+06 & $\pm$ & 2.0e+06 &   1.1 & $\pm$ & 0.3 &  26.4 \\
	S33  &   3.9e+06 & $\pm$ & 1.5e+06 &   2.1 & $\pm$ & 0.8 &  37.0 &   1.2e+07 & $\pm$ & 4.6e+06 &   0.7 & $\pm$ & 0.2 &  37.0 \\
	S38  &   2.4e+05 & $\pm$ & 6.1e+04 &  34.4 & $\pm$ & 8.7 &  25.4 &   7.6e+05 & $\pm$ & 1.9e+05 &  10.9 & $\pm$ & 2.8 &  25.4 \\
	S39  &   5.6e+05 & $\pm$ & 1.6e+05 &  14.8 & $\pm$ & 4.3 &  28.8 &   1.8e+06 & $\pm$ & 5.1e+05 &   4.7 & $\pm$ & 1.4 &  28.8 \\
	S42  &   7.3e+06 & $\pm$ & 3.8e+06 &   1.1 & $\pm$ & 0.6 &  52.7 &   2.3e+07 & $\pm$ & 1.2e+07 &   0.4 & $\pm$ & 0.2 &  52.7 \\
	S54  &   4.2e+06 & $\pm$ & 4.9e+06 &   2.0 & $\pm$ & 2.3 & 118.3 &   1.3e+07 & $\pm$ & 1.6e+07 &   0.6 & $\pm$ & 0.7 & 118.3 \\
	S55  &   2.1e+05 & $\pm$ & 5.9e+04 &  38.7 & $\pm$ & 10.7 &  27.6 &   6.8e+05 & $\pm$ & 1.9e+05 &  12.2 & $\pm$ & 3.4 &  27.6 \\
	S60  &   1.5e+06 & $\pm$ & 4.1e+05 &   5.6 & $\pm$ & 1.6 &  27.7 &   4.6e+06 & $\pm$ & 1.3e+06 &   1.8 & $\pm$ & 0.5 &  27.7 \\
	S66  &   1.9e+07 & $\pm$ & 6.0e+06 &   0.4 & $\pm$ & 0.1 &  31.4 &   6.0e+07 & $\pm$ & 1.9e+07 &   0.1 & $\pm$ & 0.0 &  31.4 \\
	S67  &   1.2e+07 & $\pm$ & 3.5e+06 &   0.7 & $\pm$ & 0.2 &  30.1 &   3.7e+07 & $\pm$ & 1.1e+07 &   0.2 & $\pm$ & 0.1 &  30.1 \\
	S71  &   2.9e+06 & $\pm$ & 1.1e+06 &   2.9 & $\pm$ & 1.1 &  37.3 &   9.2e+06 & $\pm$ & 3.4e+06 &   0.9 & $\pm$ & 0.3 &  37.3 \\
	S83  &   1.7e+07 & $\pm$ & 6.9e+06 &   0.5 & $\pm$ & 0.2 &  40.3 &   5.4e+07 & $\pm$ & 2.2e+07 &   0.2 & $\pm$ & 0.1 &  40.3 \\
	S85  &   5.1e+07 & $\pm$ & 7.2e+07 &   0.2 & $\pm$ & 0.2 & 141.0 &   1.6e+08 & $\pm$ & 2.3e+08 &   0.1 & $\pm$ & 0.1 & 141.0 \\
	S87  &   4.6e+07 & $\pm$ & 1.5e+07 &   0.2 & $\pm$ & 0.1 &  32.4 &   1.4e+08 & $\pm$ & 4.7e+07 &   0.1 & $\pm$ & 0.0 &  32.4 \\
	S89  &   7.9e+06 & $\pm$ & 3.0e+06 &   1.0 & $\pm$ & 0.4 &  37.8 &   2.5e+07 & $\pm$ & 9.5e+06 &   0.3 & $\pm$ & 0.1 &  37.8 \\
	S91  &   2.6e+07 & $\pm$ & 8.2e+06 &   0.3 & $\pm$ & 0.1 &  31.9 &   8.2e+07 & $\pm$ & 2.6e+07 &   0.1 & $\pm$ & 0.0 &  31.9 \\
	S96  &   1.9e+07 & $\pm$ & 5.6e+06 &   0.4 & $\pm$ & 0.1 &  30.0 &   5.9e+07 & $\pm$ & 1.8e+07 &   0.1 & $\pm$ & 0.0 &  30.0 \\
	S97  &   3.3e+07 & $\pm$ & 1.9e+07 &   0.2 & $\pm$ & 0.1 &  55.9 &   1.1e+08 & $\pm$ & 5.9e+07 &   0.1 & $\pm$ & 0.0 &  55.9 \\
	S145 &   1.0e+07 & $\pm$ & 6.6e+06 &   0.8 & $\pm$ & 0.6 &  66.7 &   3.1e+07 & $\pm$ & 2.1e+07 &   0.3 & $\pm$ & 0.2 &  66.7 \\
	S175 &   1.8e+05 & $\pm$ & 7.4e+04 &  45.3 & $\pm$ & 18.3 &  40.3 &   5.8e+05 & $\pm$ & 2.3e+05 &  14.3 & $\pm$ & 5.8 &  40.3 \\
	R34  &   1.7e+07 & $\pm$ & 8.6e+06 &   0.5 & $\pm$ & 0.2 &  50.5 &   5.4e+07 & $\pm$ & 2.7e+07 &   0.2 & $\pm$ & 0.1 &  50.5 \\
	R44  &   7.5e+07 & $\pm$ & 6.5e+07 &   0.1 & $\pm$ & 0.1 &  86.2 &   2.4e+08 & $\pm$ & 2.0e+08 &   0.0 & $\pm$ & 0.0 &  86.2 \\
\hline
\end{tabular}
\end{table*}

\begin{table*}
	\centering
	\caption{The same as in Table \ref{tab1} but for $f_{SP}=1.16\pm0.16$ and $f_{SP}=1.144\pm0.144$.}
	\label{tab3}
	\setlength{\tabcolsep}{0.11cm}
	\begin{tabular}{|l|rcl|rcl|r|rcl|rcl|r|}
		\hline
		&\multicolumn{7}{c|}{$f_{SP}=1.16\pm0.16$}&\multicolumn{7}{c|}{$f_{SP}=1.144\pm0.144$}\\
		\cline{2-15}
		\multicolumn{1}{|l|}{Star}&\multicolumn{3}{c|}{$\Lambda\pm\Delta\Lambda$}&\multicolumn{3}{c|}{$m_g\pm\Delta m_g$}&\multicolumn{1}{c|}{R.E.}&\multicolumn{3}{c|}{$\Lambda\pm\Delta\Lambda$}&\multicolumn{3}{c|}{$m_g\pm\Delta m_g$}&\multicolumn{1}{c|}{R.E.} \\
		&\multicolumn{3}{c|}{(AU)}&\multicolumn{3}{c|}{$(10^{-24}\ \mathrm{eV})$}&\multicolumn{1}{c|}{(\%)}&\multicolumn{3}{c|}{(AU)}&\multicolumn{3}{c|}{$(10^{-24}\ \mathrm{eV})$}&\multicolumn{1}{c|}{(\%)} \\
		\hline
		\hline
		S1   &   9.1e+05 & $\pm$ & 5.1e+05 &   9.1 & $\pm$ & 5.1 &  55.7 &   9.6e+05 & $\pm$ & 5.4e+05 &   8.6 & $\pm$ & 4.8 &  55.7 \\
		S2   &   3.7e+04 & $\pm$ & 1.9e+04 & 223.4 & $\pm$ & 114.6 &  51.3 &   3.9e+04 & $\pm$ & 2.0e+04 & 211.9 & $\pm$ & 108.7 &  51.3 \\
		S4   &   4.9e+05 & $\pm$ & 2.5e+05 &  16.8 & $\pm$ & 8.7 &  51.7 &   5.2e+05 & $\pm$ & 2.7e+05 &  15.9 & $\pm$ & 8.2 &  51.7 \\
		S6   &   5.6e+05 & $\pm$ & 2.8e+05 &  14.9 & $\pm$ & 7.5 &  50.2 &   5.9e+05 & $\pm$ & 2.9e+05 &  14.1 & $\pm$ & 7.1 &  50.2 \\
		S8   &   3.1e+05 & $\pm$ & 1.6e+05 &  26.8 & $\pm$ & 14.2 &  53.0 &   3.3e+05 & $\pm$ & 1.7e+05 &  25.4 & $\pm$ & 13.4 &  53.0 \\
		S9   &   2.5e+05 & $\pm$ & 1.4e+05 &  33.3 & $\pm$ & 18.2 &  54.7 &   2.6e+05 & $\pm$ & 1.4e+05 &  31.6 & $\pm$ & 17.3 &  54.7 \\
		S12  &   1.3e+05 & $\pm$ & 6.8e+04 &  62.4 & $\pm$ & 32.1 &  51.4 &   1.4e+05 & $\pm$ & 7.2e+04 &  59.2 & $\pm$ & 30.4 &  51.4 \\
		S13  &   3.1e+05 & $\pm$ & 1.5e+05 &  27.1 & $\pm$ & 13.7 &  50.5 &   3.2e+05 & $\pm$ & 1.6e+05 &  25.7 & $\pm$ & 13.0 &  50.5 \\
		S14  &   4.1e+04 & $\pm$ & 2.5e+04 & 204.1 & $\pm$ & 127.2 &  62.3 &   4.3e+04 & $\pm$ & 2.7e+04 & 193.6 & $\pm$ & 120.7 &  62.3 \\
		S17  &   4.9e+05 & $\pm$ & 2.5e+05 &  17.0 & $\pm$ & 8.8 &  52.1 &   5.1e+05 & $\pm$ & 2.7e+05 &  16.1 & $\pm$ & 8.4 &  52.1 \\
		S18  &   2.5e+05 & $\pm$ & 1.3e+05 &  32.9 & $\pm$ & 17.0 &  51.5 &   2.7e+05 & $\pm$ & 1.4e+05 &  31.2 & $\pm$ & 16.1 &  51.5 \\
		S19  &   5.3e+05 & $\pm$ & 3.8e+05 &  15.7 & $\pm$ & 11.2 &  71.4 &   5.5e+05 & $\pm$ & 4.0e+05 &  14.9 & $\pm$ & 10.7 &  71.4 \\
		S21  &   1.4e+05 & $\pm$ & 7.6e+04 &  59.6 & $\pm$ & 32.6 &  54.6 &   1.5e+05 & $\pm$ & 8.0e+04 &  56.6 & $\pm$ & 30.9 &  54.6 \\
		S22  &   3.3e+06 & $\pm$ & 2.3e+06 &   2.5 & $\pm$ & 1.7 &  69.1 &   3.5e+06 & $\pm$ & 2.4e+06 &   2.4 & $\pm$ & 1.6 &  69.1 \\
		S23  &   2.5e+05 & $\pm$ & 1.8e+05 &  33.1 & $\pm$ & 23.4 &  70.6 &   2.6e+05 & $\pm$ & 1.9e+05 &  31.4 & $\pm$ & 22.2 &  70.6 \\
		S24  &   7.1e+05 & $\pm$ & 4.1e+05 &  11.8 & $\pm$ & 6.8 &  58.2 &   7.4e+05 & $\pm$ & 4.3e+05 &  11.1 & $\pm$ & 6.5 &  58.2 \\
		S29  &   4.2e+05 & $\pm$ & 2.7e+05 &  19.9 & $\pm$ & 12.8 &  64.1 &   4.4e+05 & $\pm$ & 2.8e+05 &  18.9 & $\pm$ & 12.1 &  64.1 \\
		S31  &   6.0e+05 & $\pm$ & 3.1e+05 &  13.9 & $\pm$ & 7.1 &  51.4 &   6.3e+05 & $\pm$ & 3.2e+05 &  13.2 & $\pm$ & 6.8 &  51.4 \\
		S33  &   9.8e+05 & $\pm$ & 6.1e+05 &   8.4 & $\pm$ & 5.2 &  62.0 &   1.0e+06 & $\pm$ & 6.4e+05 &   8.0 & $\pm$ & 4.9 &  62.0 \\
		S38  &   6.0e+04 & $\pm$ & 3.0e+04 & 137.6 & $\pm$ & 69.3 &  50.4 &   6.4e+04 & $\pm$ & 3.2e+04 & 130.5 & $\pm$ & 65.7 &  50.4 \\
		S39  &   1.4e+05 & $\pm$ & 7.5e+04 &  59.4 & $\pm$ & 32.0 &  53.8 &   1.5e+05 & $\pm$ & 7.9e+04 &  56.3 & $\pm$ & 30.3 &  53.8 \\
		S42  &   1.8e+06 & $\pm$ & 1.4e+06 &   4.6 & $\pm$ & 3.5 &  77.7 &   1.9e+06 & $\pm$ & 1.5e+06 &   4.3 & $\pm$ & 3.4 &  77.7 \\
		S54  &   1.0e+06 & $\pm$ & 1.5e+06 &   7.9 & $\pm$ & 11.4 & 143.3 &   1.1e+06 & $\pm$ & 1.6e+06 &   7.5 & $\pm$ & 10.8 & 143.3 \\
		S55  &   5.4e+04 & $\pm$ & 2.8e+04 & 154.8 & $\pm$ & 81.4 &  52.6 &   5.6e+04 & $\pm$ & 3.0e+04 & 146.9 & $\pm$ & 77.2 &  52.6 \\
		S60  &   3.7e+05 & $\pm$ & 1.9e+05 &  22.6 & $\pm$ & 11.9 &  52.7 &   3.9e+05 & $\pm$ & 2.0e+05 &  21.4 & $\pm$ & 11.3 &  52.7 \\
		S66  &   4.8e+06 & $\pm$ & 2.7e+06 &   1.7 & $\pm$ & 1.0 &  56.4 &   5.0e+06 & $\pm$ & 2.8e+06 &   1.7 & $\pm$ & 0.9 &  56.4 \\
		S67  &   2.9e+06 & $\pm$ & 1.6e+06 &   2.8 & $\pm$ & 1.6 &  55.1 &   3.1e+06 & $\pm$ & 1.7e+06 &   2.7 & $\pm$ & 1.5 &  55.1 \\
		S71  &   7.3e+05 & $\pm$ & 4.5e+05 &  11.4 & $\pm$ & 7.1 &  62.3 &   7.7e+05 & $\pm$ & 4.8e+05 &  10.8 & $\pm$ & 6.7 &  62.3 \\
		S83  &   4.3e+06 & $\pm$ & 2.8e+06 &   1.9 & $\pm$ & 1.3 &  65.3 &   4.5e+06 & $\pm$ & 2.9e+06 &   1.8 & $\pm$ & 1.2 &  65.3 \\
		S85  &   1.3e+07 & $\pm$ & 2.1e+07 &   0.6 & $\pm$ & 1.1 & 166.0 &   1.4e+07 & $\pm$ & 2.2e+07 &   0.6 & $\pm$ & 1.0 & 166.0 \\
		S87  &   1.1e+07 & $\pm$ & 6.6e+06 &   0.7 & $\pm$ & 0.4 &  57.4 &   1.2e+07 & $\pm$ & 6.9e+06 &   0.7 & $\pm$ & 0.4 &  57.4 \\
		S89  &   2.0e+06 & $\pm$ & 1.2e+06 &   4.2 & $\pm$ & 2.6 &  62.8 &   2.1e+06 & $\pm$ & 1.3e+06 &   4.0 & $\pm$ & 2.5 &  62.8 \\
		S91  &   6.5e+06 & $\pm$ & 3.7e+06 &   1.3 & $\pm$ & 0.7 &  56.9 &   6.8e+06 & $\pm$ & 3.9e+06 &   1.2 & $\pm$ & 0.7 &  56.9 \\
		S96  &   4.7e+06 & $\pm$ & 2.6e+06 &   1.8 & $\pm$ & 1.0 &  55.0 &   4.9e+06 & $\pm$ & 2.7e+06 &   1.7 & $\pm$ & 0.9 &  55.0 \\
		S97  &   8.3e+06 & $\pm$ & 6.8e+06 &   1.0 & $\pm$ & 0.8 &  80.9 &   8.8e+06 & $\pm$ & 7.1e+06 &   0.9 & $\pm$ & 0.8 &  80.9 \\
		S145 &   2.5e+06 & $\pm$ & 2.3e+06 &   3.3 & $\pm$ & 3.1 &  91.7 &   2.6e+06 & $\pm$ & 2.4e+06 &   3.2 & $\pm$ & 2.9 &  91.7 \\
		S175 &   4.6e+04 & $\pm$ & 3.0e+04 & 181.3 & $\pm$ & 118.4 &  65.3 &   4.8e+04 & $\pm$ & 3.1e+04 & 172.0 & $\pm$ & 112.3 &  65.3 \\
		R34  &   4.3e+06 & $\pm$ & 3.2e+06 &   1.9 & $\pm$ & 1.5 &  75.5 &   4.5e+06 & $\pm$ & 3.4e+06 &   1.8 & $\pm$ & 1.4 &  75.5 \\
		R44  &   1.9e+07 & $\pm$ & 2.1e+07 &   0.4 & $\pm$ & 0.5 & 111.2 &   2.0e+07 & $\pm$ & 2.2e+07 &   0.4 & $\pm$ & 0.5 & 111.2 \\
		\hline
	\end{tabular}
\end{table*}

We also used the expressions (\ref{lambda}), (\ref{relerr}) and (\ref{gravmass}) in order to estimate the values for the range of Yukawa interaction $\Lambda$, graviton mass $m_g$, as well as their relative and absolute errors for all S-stars from Table 3 in \cite{gill17}, except for S111. The obtained estimates are presented in Tables \ref{tab1}, \ref{tab2} and \ref{tab3} for different values of $f_{SP}$.
In order to see what happens with the upper limit of the graviton mass $m_g$ when $f_{SP}$ approaching 1, and $\Delta f_{SP}$ decreases, we obtained estimates for $f_{SP}$: $f_{SP}=1.10\pm0.19$ (see left part of Table \ref{tab1}) and $f_{SP}=1.01\pm0.005$ and $f_{SP}=1.001\pm0.0005$ (see Table \ref{tab2}). We suppose that $\Delta f_{SP}$ is half of last significant digit. We have to stress that obtained estimates presented in Tables \ref{tab1} and \ref{tab2} are calculated for the same $f_{SP}$ value for all S-stars, and currently, the Schwarzschild precession is detected only for S2 by GRAVITY collaboration \cite{abut20,abut22}. In Table \ref{tab1} we used the value of $f_{SP}=1.10\pm0.19$ in order to find the current bounds on graviton mass.By comparing these results with our previous corresponding estimates from Table 2 in \cite{zakh18a}, it can be seen that the upper bound for graviton mass $m_g$ could be improved approximately in 2--3 times in the case of the current GRAVITY estimate of $f_{SP}$. The values of $m_g$ obtained in this paper from Table \ref{tab1} ($f_{SP}=1.10$) are approximately in 2--3 times smaller than from Table 2 in paper \cite{zakh18a} for S2 star. But at the same time, the errors of these estimates are very high. This is mainly caused by the fact that the current GRAVITY estimates of $f_{SP}=1.10$ and $\Delta f_{SP}=\pm 0.19$ give very high contribution of 95\% to the relative error in the last term in r.h.s. of (\ref{relerr}). Also, additional contributions (but much smaller ones) are from observations of $\Delta P$ and $\Delta e$ (see Table 3 in \cite{gill17}). As the current GRAVITY estimates of $f_{SP}=1.10$ and $\Delta f_{SP}=\pm 0.19$, we can conclude that relative error obtain in Table \ref{tab1} for $\Delta m_g / m_g$ in some cases are above 100\%, but in majority cases it is below 100\%. In case of S2 star it is 96.3\%. It means that the more precise future observations are needed for improvement of upper graviton mass bounds. That is why we also theoretically study case when $f_{SP}$ is much closer to 1 and when absolute error $\Delta f_{SP}$ is much smaller. In Table \ref{tab2} the values of $f_{SP}=1.01\pm0.005$ and $f_{SP}=1.001\pm0.0005$ are adopted with aim to give a prediction for possible improvements of $m_g$ and its uncertainty in the case if in future, more precise measurements will show that $f_{SP}$ will be much closer to 1 and $\Delta f_{SP}$ will be much smaller then $\Delta f_{SP}=\pm 0.19$. As it can be seen from Table \ref{tab2}, the possible improvements in this case for $m_g$ are even larger and could go up to $\sim 15$ times with respect to our previous constraints from Table 2 in \cite{zakh18a}, i.e. the values of $m_g$ obtained in this paper from Table \ref{tab2} are $\sim 15$ times smaller than from Table 2 in paper \cite{zakh18a} for the corresponding S-star. In this case, the last term in r.h.s. of (\ref{relerr}) gives a more acceptable contribution of 25\% to the relative error in the last term in r.h.s. of (\ref{relerr}).
From Tables \ref{tab1} and \ref{tab2} we can see that relative error $\Delta m_g / m_g$ in some cases are very high, especially for $f_{SP}=1.10\pm0.19$. If we adopt the Yukawa gravity model we obtain that $m_g$ must be in the interval $[m_g - \Delta m_g,  m_g + \Delta m_g]$. If we take for upper graviton mass bounds values $m_g + \Delta m_g$ instead of $m_g$, still it is reduced with respect to our previous constraints from Table 2 in \cite{zakh18a}.

The best fit values of $f_{SP}$ obtained in paper \cite{abut22} are smaller than 1, which would be inconvenient for the analysis (because of the factor $\sqrt{f_{SP}-1}$ in Eq. (\ref{lambda})). The best fit value is probably not always the ideal basis for analysis, since it depends not only on the data, but also on the details of the applied systematics and data analysis \cite{abut20,abut22}. According to the paper \cite{abut20} best fit value of $f_{SP}$ varies between 0.9 and 1.2 but the key observation is that all three measurements are 1 $\sigma$ compatible with the GR value 1. Also, this 1 $\sigma$ range is decreasing with the abundance and quality of the data. That is why we took the following strategy: in our analysis and calculations we assumed that the above measurements are a confirmation of GR within 1 $\sigma$ (these measurements are 1 $\sigma$ compatible with the GR value 1), and thus we started from the GR value of 1, and then added the above 1 $\sigma$ error(s), e.g. we use 1 + 0.19 = 1.19; 1 + 0.16 = 1.16 and 1 + 0.144 = 1.144 as the basis for our analysis. It means that we will take upper bound for $f_{SP}$, i.e. $f_{SP} = 1.19\pm0.19$, $f_{SP} = 1.16\pm0.16$ and $f_{SP} = 1.144\pm0.144$. In that way, we obtained estimates for the following values of $f_{SP}$: $f_{SP}=1.10\pm0.19$ from \cite{abut20} and $f_{SP} = 0.85\pm0.16$ from \cite{abut22}, both via S2, as well as $f_{SP} = 0.997\pm0.144$ from \cite{abut22} via S2, S29, S38, S55 (see the corresponding cases in Tables \ref{tab1} and \ref{tab3}).

In Table \ref{tab1} we presented results for (left part) $f_{SP} = 1.10\pm0.19$ (best fit value from \cite{abut20}) and for (right part) $f_{SP} = 1.19\pm0.19$ (instead of $f_{SP} = 1.10\pm0.19$ we use $f_{SP} = 1.19\pm0.19$) in the case of all S-stars from Table 3 in \cite{gill17} except of S111. In the second case we  started from the GR value of 1, and then added the above 1 $\sigma$ error(s), e.g. we use 1 + 0.19 = 1.19 for $f_{SP}$. It can be seen in both way of calculation, the upper bounds for graviton mass $m_g$ are very similar, in first case $m_g$ is little lower, but error is higher. For example, values of $m_g$ for S2 star are:
\begin{itemize}[nosep]
	\item[] $f_{SP} = 1.10\pm0.19$; \quad $m_g < (177\pm170) \times 10^{-24}~$eV
	\item[] $f_{SP} = 1.19\pm0.19$; \quad $m_g < (244\pm125) \times 10^{-24}~$eV.
\end{itemize}

By comparing mutually results from (Tables \ref{tab1} and \ref{tab3}) it can be seen that the upper bounds for graviton mass $m_g$ are also very similar. For example, values of $m_g$ for S2 star (instead of $f_{SP} = 0.85\pm0.16$ we use $f_{SP} = 1.16\pm0.16$; and instead of $f_{SP} = 0.997\pm0.144$ we use $f_{SP} = 1.144\pm0.144$) in case of the following values of $f_{SP}$ \cite{abut20,abut22} are:
\begin{itemize}[nosep]
\item[] $f_{SP} = 1.10\pm0.19$; \quad $m_g < (244\pm125) \times 10^{-24}~$eV
\item[] $f_{SP} = 0.85\pm0.16$; \quad $m_g < (224\pm115) \times 10^{-24}~$eV
\item[] $f_{SP} = 0.997\pm0.144$; \quad $m_g < (212\pm109) \times 10^{-24}~$eV.
\end{itemize}

Also, by comparing estimates from Table \ref{tab3} with our previous corresponding estimates from Table 2 in Ref. \cite{zakh18a} ($m_g < (548\pm32)\times 10^{-24}~$eV), it can be seen that the upper bound for graviton mass $m_g$ could be improved for about 2-3 times. In case of S2 star the relative error is 51.3\%. It means that the more precise future observations are needed for improvement of upper graviton mass bounds.

One of the biggest factors impacting the best fit result on $\sigma$ is the systematics and choice of priors for the localisation and proper motion of Sgr A*. This is mentioned in papers \cite{abut20,abut22}. The underlying systematics is studied in detail in the following papers \cite{plew15,saka19,darl23} and references therein.

\clearpage

\section{Conclusions}

Here we study whether the factor $f_{SP}$, which was recently introduced and measured by the GRAVITY Collaboration and which parametrizes the effect of the Schwarzschild metric, could be used to improve our previous constraints on the range of Yukawa gravity interaction $\Lambda$ and graviton mass $m_g$. For that purpose, we derived the relation between $\Lambda$ and the factor $f_{SP}$ and used it to study the orbits of S-stars, obtained by two different modified PPN equation of motion. The main results of the present study can be summarized as follows:
\begin{itemize}
\item The obtained relation between the range of Yukawa gravity interaction $\Lambda$ and the factor $f_{SP}$ can be used to improve the constraints on the range of Yukawa gravity interaction $\Lambda$ and graviton mass $m_g$ using the measured value of $f_{SP}$.
\item Current GRAVITY estimates of $f_{SP}=1.10\pm 0.19$ (from \cite{abut20}) and $f_{SP}=0.85\pm0.16$ and $f_{SP}=0.997\pm0.144$ (from \cite{abut22}) can improve our previous constraints on the upper bound of graviton mass for about 2-3 times (by comparing these results with our previous corresponding estimates from Table 2 in \cite{zakh18a} for the corresponding S-star) but at the same time, it results with high contribution to the relative error in the last term in r.h.s. of (\ref{relerr}). We can conclude that total relative error obtain in Table \ref{tab1} for $\Delta m_g / m_g$ in some cases are above 100\%, but in majority cases it is below 100\%. In case of S2 star it is 96.3\%. It means that the more precise future observations are needed.
\item If the future high precision observations will confirm GR prediction for Schwarzschild precession with factor around $f_{SP}=1.01$, it could improve significantly the constraints on the upper bound of graviton mass, or in other words, these estimates in $\sim 4$ times better than our previous constraints from Table 2 in \cite{zakh18a}. The possible improvements in this case for $f_{SP}=1.001$ are even larger and could go up to $\sim 15$ times with respect to our previous constraints from Table 2 in \cite{zakh18a}, but it will hold only under assumption that the same (or very similar) accuracy of measurement of $f_{SP}$ will be achieved for all the S-stars in Table \ref{tab1}. 
\item According to our theoretical results, the smaller values of the graviton mass $m_g$ are obtained for the S-stars with larger orbital periods $P$ and lower eccentricities $e$, but at the same time the resulting precession becomes smaller and thus harder to observe and larger value of period $P$ requires longer monitoring. We have to note that above claim is correct only under the assumption that $f_{SP}$ has been measured for such an orbit to a given precision.
\item If we compare results in Table \ref{tab1} with those in Table \ref{tab3} it can be seen that the upper bound for graviton mass of $m_g$ are very similar. By comparing estimates from Table \ref{tab3} with our previous corresponding estimates from Table 2 in \cite{zakh18a} ($m_g < (548\pm32)\times 10^{-24}~$eV), it can be seen that the upper bound for graviton mass $m_g$ could be improved for about 2-3 times. In case of S2 star the relative error is 51.3\%. Therefore, the more precise future observations are needed in order to improve value of upper graviton mass bounds. 
\item We compare simulated orbits obtained using Eq.(\ref{ppneom1}) and Eq.(\ref{ppneom2}) and show that these two approaches give very similar results, both models produce the same $\Delta\omega$ per orbit, but corresponding $\omega(t)$ is little different. The biggest difference among studying models are at pericenter. The difference is smaller when $f_{SP}$ is closer to the GR case. 

\end{itemize}

\clearpage

\begin{acknowledgments}
This work is supported by Ministry of Science, Technological Development and Innovations of the Republic of Serbia through the Project contracts No. 451-03-66/2024-03/200002 and 451-03-66/2024-03/200017. The authors appreciate the referee for a constructive criticism and useful remarks which helped us to revise our manuscript.

\end{acknowledgments}

\end{document}